\documentclass[11pt,onecolumn,english,showpacs,preprintnumbers,amsmath,amssymb,floatfix]{revtex4-1}

\usepackage[T1]{fontenc}
\usepackage[latin9]{inputenc}
\usepackage{color}
\usepackage{array}
\usepackage{amstext}
\usepackage{graphicx}
\usepackage{esint}
\usepackage{rotating}
\usepackage[font=small,labelfont=bf]{caption}

\usepackage[colorlinks = true,
linkcolor = blue,
urlcolor  = magenta,
citecolor = red,
anchorcolor = blue]{hyperref}

\makeatletter

\providecommand{\tabularnewline}{\\}

\@ifundefined{textcolor}{}
{%
 \definecolor{BLACK}{gray}{0}
 \definecolor{WHITE}{gray}{1}
 \definecolor{RED}{rgb}{1,0,0}
 \definecolor{GREEN}{rgb}{0,1,0}
 \definecolor{BLUE}{rgb}{0,0,1}
 \definecolor{CYAN}{cmyk}{1,0,0,0}
 \definecolor{MAGENTA}{cmyk}{0,1,0,0}
 \definecolor{YELLOW}{cmyk}{0,0,1,0}
 }

\@ifundefined{definecolor}
 {\usepackage{color}}{}
\@ifundefined{definecolor}
 {\usepackage{color}}{}
\makeatother
\usepackage{babel}

\begin{document}


\title { Phenomenology of leading nucleon production in $ep$ collisions at HERA in the framework of fracture functions }

\author{Samira Shoeibi$^{1}$}
\email{Samira.Shoeibimohsenabadi@mail.um.ac.ir}

\author{F. Taghavi-Shahri$^{1}$}
\email{Taghavishahri@um.ac.ir}

\author{Hamzeh Khanpour$^{2,3}$}
\email{Hamzeh.Khanpour@mail.ipm.ir}

\author{Kurosh Javidan$^{1}$}
\email{Javidan@um.ac.ir}

\affiliation {
$^{(1)}$Department of Physics, Ferdowsi University of Mashhad, P.O.Box 1436, Mashhad, Iran              \\
$^{(2)}$Department of Physics, University of Science and Technology of Mazandaran, P.O.Box 48518-78195, Behshahr, Iran         \\
$^{(3)}$School of Particles and Accelerators, Institute for Research in Fundamental Sciences (IPM), P.O.Box 19395-5531, Tehran, Iran    }

\date{\today}

%
%
\begin{abstract}\label{abstract}

In recent years, several experiments at the $e^-p$ collider HERA have collected high precision deep inelastic scattering (DIS) data on the spectrum of leading nucleon carrying a large fraction of the proton's energy. In this paper, we have analyzed recent experimental data on the production of forward proton and neutron in DIS at HERA in the framework of a perturbative QCD.
We propose a technique based on the fractures functions framework, and extract the nucleon fracture functions (nucleon FFs) ${\cal M}_2^{(n/p)} (x, Q^2; x_L)$ from global QCD analysis of DIS data measured by ZEUS collaboration at HERA. We have shown that an approach based on the fracture functions formalism allows us phenomenologically parametrize the nucleon FFs.
 Considering both leading neutron as well as leading proton production data at HERA, we present the results for the separate parton
distributions for all parton species, including valence quark densities, the anti-quark densities, the strange sea distribution, and the gluon distribution functions.
We proposed several parameterizations for the nucleon FFs and open the possibility of these asymmetries.
 The obtained optimum set of nucleon FFs  is accompanied by Hessian uncertainty sets which allow one to propagate uncertainties to other observables interest.
The extracted results for the $t$-integrated leading neutron $F_2^{\rm LN(3)} (x, Q^2; x_L)$ and leading proton $F_2^{\rm LP(3)} (x, Q^2; x_L)$ structure functions are in good agreement with all data analyzed, for a wide range of fractional momentum variable $x$ as well as the longitudinal momentum fraction $x_L$.

\end{abstract}

\pacs{12.38.Bx, 12.39.-x, 14.65.Bt}

\maketitle

\tableofcontents{}

%
\section{Introduction}\label{sec:Introduction}

These days, a complete understanding of the nucleon and nuclear structures in terms of the underlying partonic and nucleonic constituents is one of the outstanding challenges
in hadron physics. High energy lepton-proton and proton-proton scattering provide the most powerful tools to investigate the hadron structures.
In such processes, contributions to the measured nucleon $F_2$ and nuclear $F^A$ structure functions can be expressed in terms of the parton distribution functions (PDFs), nuclear PDFs and spin-dependent PDFs of the nucleon. Precise understanding of PDFs will be a key ingredient in searches for new physics at the LHC through, for example, top-quark and Higgs-boson coupling measurements~\cite{Gao:2017yyd,South:2016cmx}. In consequence, reliable extraction of information on the unpolarized PDFs~\cite{Ball:2017nwa,Bourrely:2015kla,Harland-Lang:2014zoa,Hou:2017khm,MoosaviNejad:2016ebo,Khanpour:2016uxh,Alekhin:2017kpj}, helicity-dependent PDFs~\cite{Khanpour:2017cha,Shahri:2016uzl,Jimenez-Delgado:2014xza,Sato:2016tuz,Leader:2014uua,Khanpour:2017fey,Nocera:2014gqa,Ethier:2017zbq} and global nuclear PDFs fitting efforts~\cite{Khanpour:2016pph,Eskola:2016oht,Kovarik:2015cma,Wang:2016mzo} from global QCD analyses of DIS data as well as all related studies~\cite{Bertone:2017tyb,Goharipour:2017rjl,Dahiya:2016wjf,Ball:2016spl,Goharipour:2017uic,Haider:2016zrk,Accardi:2016qay,Armesto:2015lrg,Frankfurt:2015cwa,Khanpour:2017slc,Salajegheh:2015xoa,Kalantarians:2017mkj,Kusina:2016fxy,Boroun:2015yea,Zarrin:2016kxf,Nocera:2017zge,Aschenauer:2017oxs,Mottaghizadeh:2017vef,Mottaghizadeh:2017eqg,Boroun:2014dka,Phukan:2017lzp,MoosaviNejad:2016qdx,AtashbarTehrani:2013qea}, provides deep understanding of the structure of hadrons in term of their quarks and gluon constituents.

HERA as a $e^\pm p$ collider and unique particle physics data sets collected by the H1 and ZEUS experiments, have provided opportunities to study high-energy electron-proton collisions beyond the electroweak scale~\cite{South:2016cmx}.
The main process in HERA which is the DIS, probes
the internal quark structure of the proton via exchanged virtual photons. The point-like nature of the virtual photon with $Q^2 \gg \Lambda_{\rm QCD}$ ensures that the photon can successfully probes the inner structure of the nucleon.
The high center-of-mass energy at HERA has provided searches for the rare processes and physics beyond the Standard Model (SM).

The advent of the HERA collider made it possible to explore a much wider region in momentum fraction $x$ and photon virtuality Q$^2$ than that
previously accessible at fixed target experiments. By applying QCD factorization~\cite{Collins:1989gx,Reya:1979zk} and employing the well-known DGLAP~\cite{Dokshitzer:1977sg,Gribov:1972ri,Lipatov:1974qm,Altarelli:1977zs} parton
evolution scheme, the HERA PDFs can extracted from the high-precision H1 and ZEUS measurements across
a large range in $x$. The extracted PDFs can be used as input to calculate predictions for the Large Hadron Collider (LHC), Large Hadron Electron Collider (LHeC)~\cite{AbelleiraFernandez:2012cc} as well as Future Circular Hadron Collider (FCC-he and FCC-hh)~\cite{Mangano:2017tke}  at much higher values of Q$^2$.

In addition to the points mentioned above, the productions of energetic neutrons and protons in electron-proton collisions have been extensively studied with the H1 and
ZEUS detectors at HERA~\cite{Chekanov:2002yh,Chekanov:2002pf,Rinaldi:2006mf,Aaron:2010ab} as well as the related phenomenological studies~\cite{Shoeibi:2017lrl,Ceccopieri:2014rpa,deFlorian:1997wi,Szczurek:1997cw,deFlorian:1998rj}.
The H1 and ZEUS experiments at HERA have studied the production of forward protons, neutrons and photons which carry a large fraction
of the longitudinal momentum of the incoming proton~\cite{Chekanov:2002yh,Chekanov:2002pf,Rinaldi:2006mf,Aaron:2010ab,Andreev:2014zka,Aaron:2011pe,Chekanov:2008tn,Chekanov:2007tv}.
These analyses have demonstrated that models of DIS are able to reproduce the forward
nucleons measurements if contributions from different production mechanisms are considered, such as one pion exchange (OPE)~\cite{Aaron:2010ab}, diffractive dissociation, elastic scattering of the proton~\cite{Aaron:2010ab,Chekanov:2008tn} as well as the fracture function formalism~\cite{Trentadue:1993ka}.
The measurements of leading nucleons also confirm the hypothesis of limiting fragmentation~\cite{Benecke:1969sh,Chou:1994dh}, according to which, in the high-energy limit, the cross section for the inclusive
production of particles in the target fragmentation regions is independent of the incident projectile energy~\cite{Chekanov:2002pf,Aaron:2010ab}.

In this paper, we have extracted the nucleon FFs ${\cal M}_2^{(n/p)} (x, Q^2; x_L)$ from global QCD analysis of DIS data at next-to-leading order (NLO). We have also shown that the fracture functions approach, works well in describing the deep inelastic leading nucleon data measured by the ZEUS collaborations at HERA. We argued that the fracture functions could open some new possibilities for studying hadron structure and open a new window to predict a variety of hard processes at hadron colliders.

The remainder of the paper is organized as follows:
We begin in Sec.~\ref{sec:Theory} by reviewing the formalism for the leading nucleon production
including a summary on the several models in which can be used to describe such processes.
We will focus on the fracture functions formalism in Sec.~\ref{sec:The-Fracture-Functions} in which our analysis is based on. The semi-inclusive cross section and the corresponding leading nucleon structure functions have been discussed in Se.~\ref{sec:leading-nucleons-structure-functions}. The singlet and gluon evolution equations for the nucleon FFs are presented in Sec.~\ref{sec:Evolution}.
The methodology underpinning our global QCD analysis is presented in Sec.~\ref{sec:method}, where we describe in details the parametrizations employed highlighting several improvements in the methodology compared to that introduced originally in the global analysis of the {\tt SKTJ17} neutron FFs~\cite{Shoeibi:2017lrl}.
The ZEUS-02~\cite{Chekanov:2002pf} leading neutron, and ZEUS-06~\cite{Rinaldi:2006mf} and ZEUS-09~\cite{Chekanov:2008tn} leading proton productions data sets analyzed in this study are summarized in Sec.~\ref{sec:Data}. The $\chi^2$ minimization as well as the treatment of uncertainties are presented in Sec.~\ref{uncertainties}. Our analysis results have been discussed in details in Sec.~\ref{sec:results}.
We compare our theory predictions with the fitted leading neutron and leading proton observables finding overall a good agreement with all data analyzed. 
Finally, in Sec.~\ref{sec:Summary}, we summarize our findings and preview future extensions of the present analysis.

%
\section{Theory setup}\label{sec:Theory}

In the framework of perturbative QCD, the study of leading nucleon production in lepton-proton scattering represents an important field of investigation. Indeed leading nucleon carry a significant fraction of the initial momentum $x_L>0.2$ and have low transverse momentum $p_T<0.7 \, {\rm GeV}$, therefore covering kinematic regions of the phase space are not accessible from other processes. However, due to the difficulty of detecting the leading particles in high energy physics experiment, the data available are scarce.

From the theoretical point of view, much successful phenomenological models have been developed to explain the leading nucleon production mechanism.
An alternative model to describe the leading nucleon production is based on the the Fracture Functions (FFs) formalism, where the leading particles production
is described in terms of structure functions of the fragmented nucleon~\cite{Trentadue:1993ka,Ceccopieri:2007th,Szczurek:1997cw}. Another picture is given by the Regge formalism, in which the leading nucleon are produced via the exchange of a particle mediating the interaction.
Regge theory~\cite{Regge:1959mz} gives a good description of soft hadronic interactions and can explain the leading nucleon production mechanism~\cite{Szczurek:1997cw}.
The leading nucleon production can be explained by the exchange of the Reggeon trajectory as well as a Pomeron exchange which produces only leading
proton and dominates at $x_L \sim 1$~\cite{Levman:2001kf,Levman:2002cn}.

Successful descriptions of the available data on the charge-exchange processes, $p \to n$, in hadron-hadron and lepton-hadron interactions has been obtained using the exchange of virtual
particles with the quantum numbers of the $\pi$ and $\rho$ mesons. In these kind of processes, the pion, due to its small mass, dominates the $p \to n$ transition amplitude.
One-Pion-Exchange (OPE) model, therefore, can describes by the process in which a leading neutron is produced.
For the leading neutrons production in DIS as well as in dijet photoproduction, the dominating
mechanism for $x_L > 0.6$ is the One-Pion-Exchange (OPE)~\cite{Aaron:2010ab}. Based on the assumption that at high $x_L$ the leading-neutron production is dominated by the pion exchange mechanism, the measurements of the $F_2^{LN(3)} (x, Q^2; x_L)$ can provide an important information on the pion structure function $F_2^{\pi^+}$~\cite{Aaron:2010ab}.

It is worth noting here that, the present global QCD analysis is based on the fracture functions approach which can provide a QCD-based description of
semi-inclusive DIS in the target fragmentation region~\cite{Trentadue:1993ka,Ceccopieri:2007th,Szczurek:1997cw}. This formalism, where the leading particles production is described in terms of structure functions of
the fragmented proton, has been successfully used to describe leading-nucleon production from the H1 and ZEUS collaborations~\cite{Shoeibi:2017lrl,Ceccopieri:2014rpa,deFlorian:1998rj,deFlorian:1997wi}.

%
\subsection{The Fracture Functions}\label{sec:The-Fracture-Functions}

The fracture functions approach~\cite{Trentadue:1993ka,Ceccopieri:2007th,Szczurek:1997cw} has been introduced to extend the usual QCD-improved parton description of semi-inclusive DIS to the low transverse momentum region of phase space, where the target fragmentation contribution becomes important.
In the QCD parton model, experimental cross-section can be computed by convoluting some uncalculable process-independent quantities with calculable process-dependent elementary cross sections.
Let us consider a semi-inclusive deep inelastic lepton hadron scattering process, $\ell + A \to \ell^{\prime} + h + H + X$.
In the final state, the hadronic system $X$ and $H$ are originated from the target fragmentation and from the hard interaction, respectively, and $h$
is a singled out hadron. This kind of process can receive contributions from two well separated kinematical regions for the produced hadron $h$. Thus one can write

\begin{eqnarray}\label{processes}
\sigma (\ell + A \to \ell^{\prime} + h + H + X) &=& \nonumber \\
&& \sigma (\ell + A \to \ell^{\prime} + (h + H) + X)    \nonumber \\
&& + \sigma (\ell + A \to \ell^{\prime} + H + (h + X))  \,.
\end{eqnarray}

The first term known as $\sigma_{\rm current}$  in which arising from target structure function, and only knowledge of the perturbatlvely uncalculable fragmentation function $D^h_i$ is needed. Such contributions have been widely discussed in the literature (see, e.g,~\cite{deFlorian:2017lwf,Bertone:2017tyb,Ethier:2017zbq,Anderle:2015lqa} for a clear review).

The second term in Eq.~\eqref{processes} which known as $\sigma_{\rm target}$ requires a new non-perturbative (but measurable) quantity, a fragmentation-structure or nucleon fracture ``functions'' (FFs). The $\sigma_{\rm target} (x_L)$ is given by~\cite{Trentadue:1993ka,Ceccopieri:2007th,Ji:2004wu,Ji:2004xq}

\begin{equation}\label{sigmatarget}
\sigma_{\rm target} (x_L) = \sum_{i} \int_{0}^{1-x_L} \frac{dx}{x} {\cal M}_i^{n/h} (x, Q^2; x_L)  \, \sigma_i^{\rm hard} (x, Q^2)  \,.
\end{equation}

Here $n$ refers to the detected leading nucleon in the final state and ${\cal M}_i^{n/h} (x, Q^2; x_L)$ are the ordinary nucleon fracture functions which cannot be computed in perturbative QCD and like PDFs one need to determine it from fit to leading nucleon data. They are expected to give the dominant contributions to the cross sections for the production of leading hadrons and they can be thought as an ingredient of the perturbative QCD treatment. They refers to the non-perturbative parton distributions of a incoming hadron fragmented into a forward nucleon. Hence, they contains information about partons inside the detected leading nucleon in the final state. $\sigma_i^{\rm hard} (x, Q^2)$ in Eq.~\eqref{sigmatarget} is the particular hard lepton-parton cross-section we are interested in. Indices \textit{i} in Eq.~\eqref{sigmatarget} stands for the active parton, which interact with the incoming virtual photon, therefore the other partons in the incoming hadron would be the spectators. As we mentioned, the kinemtical variable $x$ is the momentum fraction of struck parton, and $x_L$ is defined as the longitudinal momentum fraction in which carried by the final state leading nucleon. It it worth noting here that, $1-x_L$ is the maximum available fractional momentum of the parton participating in the hard scattering, and hence, in Eq.~\eqref{sigmatarget}, one need to consider the maximum available fractional momentum in the integration. We should stress here that the Eq.~\eqref{sigmatarget} for the semi-inclusive DIS cross section at low transverse momentum, valid up to power corrections in QCD and to all orders in perturbation theory~\cite{Grazzini:1997ih,Ji:2004wu,Ji:2004xq,Ji:2005nu}.

On the theory side, there has been great progress in the last few years for the factorization hypothesis in lepton-nucleon collisions.
There were also great progress for applying the perturbative QCD to the description of the semi-inclusive DIS,
in particular, for the QCD factorization for SIDIS process. Unlike the inclusive DIS, SIDIS is more involved because of additional hadron measurement in the final state.
For the SIDIS process, one need to integrate out the transverse momentum ($p_T$) of the final state hadrons, and similar to the case of inclusive DIS, a collinear factorization is applicable.
The cross section for the SIDIS process can be written as a convolution of the integrated PDFs and the hard partonic cross
sections which can be calculated from perturbative QCD. For the low transverse momentum hadron production, in which a
collinear factorization approach may not be applicable because the transverse momentum ($p_T$) of
the final state hadron is small compared to the hard scale quantity, $Q$, one need to introduce a new factorization theorem, involving the transverse
momentum ($p_T$) dependent PDFs. It is instructive to demonstrate this factorization, but it is beyond the scope of our analysis. We refer the reader to the detailed discussions in which can be found in Ref.~\cite{Ji:2004wu,Grazzini:1997ih}.

Eq.~\eqref{sigmatarget} states that the factorization describes the full target fragmentation in terms of the fracture functions, without separating the contributions of
the active parton in which contribute in the hard scattering interactions and that of the spectators. Consequently, the fracture functions tell about the structure functions of the target hadron once it has fragmented
into a specific final state hadron $h$. ${\cal M}_i^{n/h} (x, Q^2; x_L)$ measures the conditional probability of finding a parton $i$ with momentum fraction $x$ of the incoming hadron momentum $A$, while a hadron $h$ with longitudinal momentum fraction $x_L$ is detected in the final state. At the phenomenological level, it has been shown that the fracture functions can well produce the HERA diffractive structure functions~\cite{deFlorian:1998rj} as well as leading neutron data~\cite{Shoeibi:2017lrl,Ceccopieri:2014rpa}, thus convalidating a common perturbative QCD approach to these particular classes of semi-inclusive processes.  Hence, the fracture function approach provides an alternative tool in the framework of QCD to describe the leading nucleon production mechanism.

%
\subsection{ Leading-nucleons structure functions and observables }\label{sec:leading-nucleons-structure-functions}

In comparison to the DIS and diffractive DIS (DDIS) formalism, the cross section of leading nucleons production $\frac{d^4 \sigma^{LB(4)}}{dx dx_L dQ^2 dp^2_T}$ can be expressed in terms of leading-Baryons structure functions $F_2^{LB(4)}$~\cite{Levman:2002cn,Levman:2001kf}

\begin{equation}
\frac{d^4 \sigma^{LB(4)}}{dx dx_L dQ^2 dp^2_T}  = \frac{4 \pi \alpha^2}{x Q^4} (1 - y + \frac{y^2}{2}) F_2^{LB(4)} (x, Q^2; x_L, p^2_T) (1+ \Delta_{LB}) \,,
\end{equation}

where $x_L \sim \frac{E_B}{E_p}$ is the energy fraction carried by the produced Baryons or equivalently the longitudinal momentum fraction of the detected Baryons $(B)$ in the final state, and $\Delta_{LB}$ takes into account the effect of the longitudinal structure functions $F_L^{LB(4)}(x, Q^2; x_L, p^2_T)$. The transverse momentum of the nucleons in given by $p_T \simeq x_L E_p \theta_n = 0.656 x_L \, {\rm GeV}$ for ZEUS-02 experiment~\cite{Chekanov:2002pf}, $p_T^2 < 0.5 \, {\rm GeV}^2$ for ZEUS-06~\cite{Rinaldi:2006mf} and ZEUS-09~\cite{Chekanov:2008tn} experiments.
In the OPE approach, the so-scaled fractional momentum variable $\beta$ is defined by
\begin{equation}
\beta = \frac{x}{1-x_L} \,,
\end{equation}
and $x$ is the Bjorken scaling variable. The quantity $\beta$ may be interpreted as the fraction of the exchange object's momentum carried by the gluon or quarks interacting with the virtual photon.  In term of these variables, the squared four-momentum transfer from the target proton is given by

\begin{equation}\label{t}
t \simeq - \frac{p_T^2}{x_L} - t_0 = - \frac{p_T^2}{x_L} - \frac{(1-x_L)^2}{x_L} m_p^2 \,.
\end{equation}

In the measurements presented by H1 and ZEUS collaborations, $p^2_T$ is not measured.
The integration over measured range of $dp^2_T$ up to maximum experimentally accessible range of $\theta_n^{max}$, corresponding to a $p_T^{max}$, gives

\begin{equation}\label{intpT}
F_2^{LB(3)}(x, Q^2; x_L)  = \int_{0}^{p^{2 \, max}_T} F_2^{LB(4)} (x, Q^2; x_L, p^2_T)  dp^2_T \,.
\end{equation}

It is sometimes more convenient to discuss the measurements in term of the reduced $e^+p$ cross section $\sigma^{LB(3)}_r$ which can be written as

\begin{eqnarray}\label{eq:reduced}
\sigma_r^{LB(3)} (x, Q^2; x_L)  =  F_2^{LB(3)} (x, Q^2; x_L) - \frac{y^2}{1+(1-y)^2} F_L^{LB(3)} (x, Q^2; x_L)\,.
\end{eqnarray}

where $F_2^{\rm  LB(3)}$ is the leading baryons transverse and $F_L^{\rm  LB(3)}$ is the longitudinal structure functions~\cite{Aaron:2010ab,Ceccopieri:2014rpa}.
We have to mention here that, based on hard scattering factorization and like for the case of diffractive DIS~\cite{Collins:1997sr}, the leading-baryons structure functions can be written in terms of the fragmentation-structure or nucleon ``fracture'' function and hard-scattering coefficient functions~\cite{Grazzini:1997ih,Ceccopieri:2014rpa} as,

\begin{eqnarray}\label{eq:factorization}
F^{LB(4)} (x, Q^2; x_L, p_T^2)  = \sum_{i} \int_{x}^{1} \frac{d \xi}{\xi} {\cal M}^B_{i/p} (x, \mu_F^2; x_L, p_T^2)  \times C_i (\frac{x}{\xi}, \frac{Q^2}{\mu_F^2}, \alpha_s(\mu^2_R)) + {\cal O}  (\frac{1}{Q^2}) \,.
\end{eqnarray}

The index $i$ runs on the flavour of the interacting parton and the Wilson coefficient functions, $C_q$ and $C_g$, are the same as in fully inclusive DIS~\cite{Vermaseren:2005qc}.
The $p_T$-unintegrated leading nucleon FFs  appearing in Eq.~\eqref{eq:factorization} obey the standard DGLAP evolution equations~\cite{Ceccopieri:2014rpa}.
As one can see from Refs.~\cite{Aaron:2010ab,Adloff:1998yg,Chekanov:2002pf,Chekanov:2002yh,Chekanov:2008tn} as well as we discussed in section~\ref{sec:Theory}, in order to describe the leading Baryons production, two more variable in addition to the DIS kinemtical variables are also needed which are $x_L$ and $t=(p_p-p_B)^2$. One can conclude that the ${\cal M}_i$ in Eq.~\eqref{eq:factorization} is the $t$ or $p_T$-dependent fracture functions of Eq.~\eqref{sigmatarget}. 
In the limit of $t \ll Q^2$, the dominant mechanism is target fragmentation and this expansion holds up to corrections suppressed by
powers of $\frac{1}{Q^2}$~\cite{Grazzini:1997ih}.  	
The relation between ordinary fracture functions ${\cal M}_i^{h/A} (x, Q^2; x_L)$ and the extended fracture functions ${\cal M}^B_{i/p} (x, \mu_F^2; x_L, p_T^2)$ considering $\epsilon < 1$ can be written as~\cite{Grazzini:1997ih,Ceccopieri:2007th}

\begin{equation}
{\cal M}_i^{h/A} (x, Q^2; x_L) = \int^{\epsilon Q^2} dp_{T}^{2}  {\cal M}^B_{i/p} (x, \mu_F^2; x_L, p_T^2) \,,
\end{equation}

The above ordinary fracture function has been obtained by integral over $p_{T}$ up to a cut-off of order $Q^2$, e.g. $\epsilon Q^2$.
In the leading nucleon production, the $p_T$ of the leading nucleon is integrated up to some $p_{T,max}$. We will return to this issue in section~\ref{sec:Data}.

%
\subsection{Evolution of the nucleon FFs}\label{sec:Evolution}

As we mentioned earlier, fracture functions can provide a QCD-based description of semi-inclusive DIS in the target fragmentation regions. This approach has been used to describe the leading-nucleon production data~\cite{deFlorian:1998rj,deFlorian:1997wi} and to extract the neutron FFs~\cite{Shoeibi:2017lrl,Ceccopieri:2014rpa}.
Like for the DIS structure function, QCD can not predicted the shape of the nucleon FFs. As for the global analysis of the parton distribution functions (PDFs),
the non-perturbative nucleon FFs can be parameterized at a given initial scale $Q^2_0$. Having at hand the input functional form for the nucleon FFs and the QCD-evolution as well as the corresponding observable, one can extract the nucleon FFs from a global analysis of the DIS leading-nucleon production data. The extracted nucleon FFs can be used to describe the leading nucleon production. These kind of studies can open a new window to predict a variety of hard processes at hadron colliders in which suitable hadronic triggers are used. The presence of leading nucleons in the final state indicates to the long distances processes while the short distances are probed by high photon virtuality Q$^2$. In consequence, the study of leading nucleon productions provides information on the relationship between the soft and hard aspects of the strong interaction. These statements and the corresponding discussions can be found in Refs.~\cite{Levman:2001kf,Levman:2001fcs}.

In Refs.~\cite{Adloff:1998yg,Chekanov:2002pf,Aaron:2010ab} have been shown that the leading-neutron structure function $F_2^{LN(3)}$ and inclusive DIS structure functions $F_2^p$ have a similar $(x, Q^2)$ behavior as expected from the hypothesis of limiting fragmentation~\cite{Benecke:1969sh,Chou:1994dh} in which states that the production of of leading-neutron in the proton fragmentation region is independent of $x$ and $Q^2$. The results demonstrated that the similarity between the $Q^2$ evolution of $F_2^{LN(3)}$ and $Q^2$ evolution of $F_2^p$. Therefore, as in the inclusive case, the $Q^2$ evolution of the semi-inclusive multi-particle distributions can be predicted in perturbative QCD.
Since the scale dependence of the cross section in forward particle production in DIS can be calculated within perturbative QCD~\cite{Trentadue:1993ka}, therefore the nucleon FFs also obey the standard DGLAP evolution equations~\cite{Shoeibi:2017lrl,Ceccopieri:2014rpa,Camici:1998bg,Ceccopieri:2007th}.
The evolution equations of nucleon FFs are easily obtained by the DGLAP evolution equations~\cite{Shoeibi:2017lrl,Trentadue:1993ka,Ceccopieri:2007th} as

\begin{eqnarray}\label{eq:DGLAP}
Q^2 \frac{\partial {\cal M}^n_{\Sigma/P} (x, Q^2; x_L)}{\partial Q^2} &=& \frac{\alpha_s(Q^2)}{2 \pi}  \int_{x}^{1} \frac{du}{u} P_{\Sigma}^j(u) \, {\cal M}^n_{\Sigma/P} (\frac{x}{u},  x_L, Q^2)\,, \\
Q^2 \frac{\partial {\cal M}^n_{g/P} (x, Q^2; x_L)}{\partial Q^2}  &=& \frac{\alpha_s(Q^2)}{2 \pi}  \int_{x}^{1} \frac{du}{u} P_{g}^j(u) \, {\cal M}^n_{g/P} (\frac{x}{u},  x_L, Q^2)\,,
\end{eqnarray}

where ${\cal M}^n_{\Sigma/P} (x, x_L, Q^2)$ and ${\cal M}^n_{g/P} (x, x_L, Q^2)$ are the singlet and gluon distributions, respectively.
$P_{\Sigma}$ and $P_{g}$ are the common NLO contributions to the splitting functions which are perturbatively calculable as a power expansion in the strong coupling constant $\alpha_s$. They are the same as in the case of fully inclusive DIS~\cite{Vogt:2004mw,Moch:2004pa}.
We have parametrized these non-perturbative distributions, nucleon FFs, at an input scale $Q_0^2 < m_c^2$. Their evolution to higher scale, $Q^2 > Q_0^2$, have been described by using the DGLAP evolution equation within the zero-mass variable flavour number scheme (ZM-VFNS) at the NLO accuracy of perturbative QCD.

%
\section{Outline of the analysis}\label{sec:method}

In this section, we present the method of global QCD analysis of {\tt STKJ17} nucleon FFs. The need for well-constrained nucleon FFs has recently been
highlighted in global analyses of leading neutron production~\cite{Shoeibi:2017lrl,Ceccopieri:2014rpa}. Differences between these nucleon FFs determinations come from a variety of sources,  including different data sets used in the analyses, the choice of parametrization for the nucleon FFs, assumptions about nucleon FFs that are not well constrained by data, or even the method of minimizations. Most of the analyses to date have been performed at leading-order (LO)~\cite{deFlorian:1998rj} and NLO~\cite{Shoeibi:2017lrl,Ceccopieri:2014rpa} accuracy in the strong coupling constant.

Note that our aim in this analysis is a definitive determination of nucleon FFs and to explore the application of the fracture function approach to determine the maximal information that can be extracted from the HERA leading nucleon production processes. We will try to propose several parameterizations to consider the asymmetric parton densities in which require the inclusion of most of the possible processes that have sensitivity to nucleon FFs.

%
\subsection{Input parametrizations}\label{sec:Input}

In this section we describe {\tt STKJ17} nucleon FFs at the input scale Q$_0^2 = 2 \, {\rm GeV}^2$.
In choosing a functional form for the nucleon FFs at the input scale, it is important to note that the current leading nucleon production observables
are sensitive only to the singlet and gluon distributions. In our previous analysis we therefore seek only to extract the singlet $x {\cal M}^n_{\Sigma/P}$ and gluon $x {\cal M}^n_{g/P}$ distributions, and do not attempt to separate quark and antiquark FFs~\cite{Shoeibi:2017lrl}. This would require another kind of parametrizations as well as additional data to provide a filter on the quark and antiquark flavors.

It is worth mentioning to notice that the quark distributions for the nucleon FFs at large values of $\beta=\frac{x}{1 - x_L}$ could show valence-like structures for some quark-flavour combinations. Although, the accessible values of $\beta$ or equivalently $x$ in the analyzed experimental data on leading nucleon production are quite low.
We should emphasized again that, using hypothesis of limiting fragmentation~\cite{Benecke:1969sh,Chou:1994dh}, the structure functions for the leading Baryons production is given by~\cite{Adloff:1998yg}
\begin{eqnarray}
\label{xQ2} F^{LB(4)} (x, Q^2; x_L, p_T^2) &=& f(x_L, p_T^2) \, F^{LB(2)} (x, Q^2)\,.
\end{eqnarray}
As one can see, the first term relates to Baryon variables, \{$x_L$, $p_T$\}, and the second term related to Lepton variables, \{$x$, $Q^2$\}.
In view of this fact, and in order to account the light quark decomposition, one can assume the following general initial functional form for the singlet $x {\cal M}^N_{\Sigma/P} (x, Q_0^2; x_L)$ and gluon $x {\cal M}^N_{g/P} (x, Q_0^2; x_L)$ distributions at the input scale of Q$_0^2 = 2 \, {\rm GeV}^2$
\begin{eqnarray}\label{eq:PDFQ0-OLD}
x {\cal M}^n_{\Sigma/P} (x, Q_0^2; x_L) &=& {\cal W}_q(x_L) \, \, xf_q^{\rm GJR08}(x, Q^2)  \,, \nonumber \\
x {\cal M}^n_{g/P} (x, Q_0^2; x_L) &=& {\cal W}_g(x_L) \, \, xg^{\rm GJR08}(x, Q^2)   \,,
\end{eqnarray}
where $n$ refers to the nucleon and the singlet ${\cal W}_q(x_L)$ and gluon ${\cal W}_g(x_L)$ weight functions define as
\begin{eqnarray}\label{eq:W-OLD}
{\cal W}_q(x_L)={\cal N}_{q} \, x_L^{A_q} (1 - x_L)^{B_q} (1+C_{q} \, x_L^{D_q})  \,, \nonumber \\
{\cal W}_g(x_L)={\cal N}_{g} \, x_L^{A_g} (1 - x_L)^{B_g} (1+C_{g} \, x_L^{D_g})  \,.
\end{eqnarray}
The label of $\,\Sigma/P$ and $\,g/P$ correspond to the singlet and gluon distributions inside proton, respectively. The $x_L$ dependence of the nucleon FFs is encoded in the weight function ${\cal W}_i(x_L)$. We parametrized only the ${\cal W}_q(x_L)$ and $ {\cal W}_g(x_L)$ which are related to the Baryon variables. For the $xf_q$ and $xg(x, Q^2)$ we have used the {\tt GJR08} PDFs~\cite{Gluck:2007ck}, and hence, this method allows the consideration of the asymmetry of leading nucleon FFs. The initial scale, Q$_0^2 = 2 \, {\rm GeV}^2$, is chosen at the lowest possible value where a perturbative QCD description can be applied. $xf_q^{\rm GJR08}(x, Q^2)$ and $ xg^{\rm GJR08}(x, Q^2)$ are the GJR08 parton densities for singlet and gluon distributions~\cite{Gluck:2007ck}. We have selected {\tt GJR08} PDFs set because we have used the inclusive structure functions $F_2^p(x, Q^2)$ which are obtained from the {\tt GJR08} parameterization to account the ZEUS data for the cross section ratio $r^{LP(3)} (x, Q^2; x_L)$. This selection help us to avoid inconsistencies between the theoretical formalism and the data.

We have also consider a ``MRST-type''~\cite{Martin:2001es} or old ``GRV-type''~\cite{Gluck:1993im} instead of {\tt GJR08}, and found consistence results. We have also tested our analysis by considering an independent parametrization for the $x$ dependence of $xf_q (x, Q^2)$ and $xg (x, Q^2)$.
The functional forms in Eqs.~\eqref{eq:PDFQ0-OLD} and \eqref{eq:W-OLD} state that the dependence of the input parametrizations on the lepton variables, $x$ and $Q^2$, should be independent of the leading neutron variables $x_L$.

Considering Eqs.~\eqref{eq:PDFQ0-OLD} and \eqref{eq:W-OLD}, one can also apply quark and antiquark asymmetries and present the results for the separate parton
distributions for all parton species, including valence quark densities, the anti-quark densities, the strange sea distribution functions, and the gluon distribution.
We proposed the following parameterizations for the nucleon FFs and open the possibility of these asymmetries. {\tt STKJ17} nucleon FFs at input scale reads

\begin{eqnarray}\label{eq:PDFQ0-New}
&&	x {\cal M}^n_{u_v/P} (x, Q_0^2; x_L) = {\cal W}_{u_v}(x_L) \,\, xu_v^{\rm GJR08}(x, Q^2)  \,, \nonumber \\
&&	x {\cal M}^n_{d_v/P} (x, Q_0^2; x_L) = {\cal W}_{d_v}(x_L) \,\, xd_v^{\rm GJR08}(x, Q^2)  \,, \nonumber \\
&&	x {\cal M}^n_{\Delta/P} (x, Q_0^2; x_L) = {\cal W}_{\Delta}(x_L) \,\, x \Delta^{\rm GJR08}(x, Q^2)  \,, \nonumber \\
&&	x {\cal M}^n_{(\bar{d}+\bar{u})/P} (x, Q_0^2; x_L) = {\cal W}_{(\bar{d}+\bar{u})}(x_L) \,\, x (\bar{d}+\bar{u})^{\rm GJR08}(x, Q^2)  \,, \nonumber \\
&&	x {\cal M}^n_{(s={\bar s})/P} (x, Q_0^2; x_L) = {\cal W}_{(s={\bar s})}(x_L) \,\, x {(s={\bar s})}^{\rm GJR08}(x, Q^2)  \,, \nonumber \\	
&&	x {\cal M}^n_{g/P} (x, Q_0^2; x_L)   = {\cal W}_g(x_L) \,\, xg^{\rm GJR08}(x, Q^2)   \,, \nonumber \\
\end{eqnarray}

where $n$ refers to the nucleon and the weight functions ${\cal W}_q(x_L)$ and ${\cal W}_g(x_L)$ are now defined as

\begin{eqnarray}\label{eq:W-New}
&&	{\cal W}_{u_v}(x_L) = {\cal N}_{u_v} \,\, x_L^{A_{u_v}} (1 - x_L)^{B_{u_v}} ( 1 + C_{u_v} \, x_L^{D_{u_v}} )  \,, \nonumber \\
&&	{\cal W}_{d_v}(x_L) = {\cal N}_{d_v} \,\, x_L^{A_{d_v}} (1 - x_L)^{B_{d_v}} ( 1 + C_{d_v} \, x_L^{D_{d_v}} )  \,, \nonumber \\
&&	{\cal W}_{\Delta}(x_L) = {\cal N}_{\Delta} \,\, x_L^{A_{\Delta}} (1 - x_L)^{B_{\Delta}} ( 1 + C_{\Delta} \, x_L^{D_{\Delta}} )  \,, \nonumber \\
&&	{\cal W}_{(\bar{d}+\bar{u})}(x_L) = {\cal N}_{(\bar{d}+\bar{u})} \,\, x_L^{A_{(\bar{d}+\bar{u})}} (1 - x_L)^{B_{(\bar{d}+\bar{u})}} ( 1 + C_{(\bar{d}+\bar{u})} \, x_L^{D_{(\bar{d}+\bar{u})}} )  \,, \nonumber \\
&&	{\cal W}_{s}(x_L) = {\cal W}_{\bar s}(x_L) = \frac{1}{4} \times {\cal W}_{(\bar{d}+\bar{u})}(x_L) \,, \nonumber \\	
&&	{\cal W}_g(x_L) = {\cal N}_g \,\, x_L^{A_g} (1 - x_L)^{B_g} ( 1 + C_g \, x_L^{D_g} )  \,,
\end{eqnarray}

where $p_i = \{{\cal N}_i, A_i, B_i, C_i, D_i\}$ are the free parameters to be fitted. The total number of free parameters for all the neutron FFs is 30. However, we have to fix by hand certain shape parameters that are difficult to constrain by data in order to obtain a reasonable QCD fit. With the available leading nucleon data sets, for example, the fit can not constrain distinct strange-quark fracture functions. We therefore assume a symmetric strange-quark distributions. The {\tt GJR08} PDFs considered a symmetric strange distributions which is $xs^{\rm GJR08}(x, Q^2) = x \bar{s}^{\rm GJR08}(x, Q^2) = k (x \bar{d}^{\rm GJR08}(x, Q^2) + x \bar{u}^{\rm GJR08}(x, Q^2))$ with $k = 1/4$.  Following that, we apply ${\cal W}_{s}(x_L) = {\cal W}_{\bar s}(x_L) = \frac{1}{4} \times {\cal W}_{(\bar{d} + \bar{u})}(x_L)$  in our analysis.
Therefore, our strange densities can be obtained from this assumption and directly sensitive to the ${\bar u}$-type from ${\bar d}$-type distributions.
We should mentioned here that, one could also consider a standard ``MSTW-type'' parameterizations for the total light sea quark distribution as $xS = 2 \bar u + 2 \bar d + s + \bar s$ as~\cite{Martin:2009iq}
\begin{eqnarray}
xS = x^{a_{S}} (1-x)^{b_{S}} (1 + c_S \sqrt{x} + d_S x)\,.
\end{eqnarray}
However, in our analysis we prefer to consider the {\tt GJR08} NLO parton sets.

As can be seen from Eq.~\eqref{eq:W-New}, we consider a standard ``GJR-type'' parameterizations for the up and down valence quark distributions. Since we have used both the leading proton (LP) and leading neutron (LN) data in our analysis with a possible separation of the $u$ and $d$ densities, we believe that these data sets can provide the possibility to determination of valence quark distributions. The distributions of nucleon FFs in the quark sector at large $x$ may show valence-like structures for some quark-flavour
combinations~\cite{Ceccopieri:2014rpa}. Since the accessible values of $x$ and $x_L$ in the experimental data are quite wide, these data sets can constrain the up and down valence quark distributions well enough. The kinematical coverage of our analyzed data sets can be found in Figs.~\ref{fig:ZEUS-FLN}, \ref{fig:ZEUS-FLP} and \ref{fig:ZEUS-09Data}.
Finally, note that the nucleon FFs parametrizations adopted in this analysis is intrinsically more flexible than those used in our previous global analysis.
We will show that the above parametrizations will provide us the enough flexibility to extract the non-singlet, valence-like and gluon densities.

The effects of heavy quark masses ($m_c$, $m_b$) in hard processes and the appropriate definition of parton densities for these quarks have been very actively studied in recent years.
When heavy quarks participate in hard processes, the simplest and easiest choice is to treat these particles as massless throughout the calculation, rather than appeal to the more conventional
massive parton approximation. However, in most modern global QCD analysis of data, the so called general mass variable flavor number scheme (GM-VFNS)~\cite{Martin:2009iq} for
parton densities are used which is the actual relevance of heavy quark mass corrections. For the case of leading nucleon production at collider DIS as well as hadron collider, a little attention has been paid to the heavy quark contributions.
Our analysis in this work has been performed at NLO accuracy in perturbative QCD in the $\overline{MS}$ scheme.
For the heavy quarks $c$ and $b$ we have used the zero-mass variable flavor scheme (ZM-VFNS) and activate the
heavy quarks at their mass thresholds, $m_c = 1.43 \, {\rm GeV}$ and $m_b = 4.5 \, {\rm GeV}$. The strong coupling has been fixed on the world average value of $\alpha_s (M_Z^2) = 0.118$~\cite{Patrignani:2016xqp}.

%
\subsection{Selection of data sets}\label{sec:Data}

Measurements of the cross sections for the production of leading nucleon at very small angles
with respect to the proton beam direction in deep-inelastic electron-proton scattering at HERA have been presented as a function of the
scale variables $\beta$, $x_L$ and the photon virtuality Q$^2$.
In this section, we review the experimental results from H1 and ZEUS collaborations on the production of leading nucleon in neutral current $e^+p$ collisions.
The rate of leading nucleon production depends only logarithmically on the virtuality of exchanged photon Q$^2$ and the momentum fraction of the incoming proton carried by struck quarks, $\beta = 1/(1- x_L)$. These leading nucleon carry a large fraction of the incoming proton's energy $x_L \geq 0.2$ and produce with low transverse momentum $p_T  \leq  0.7$ GeV.
The data from HERA collider cover a large kinematic range of photon virtuality Q$^2$, and scale variables of $\beta$ and $x_L$. These leading nucleons production measurements includes both semi-inclusive DIS $e p \to e n X$~\cite{Aaron:2010ab,Chekanov:2002pf,Adloff:1998yg,Chekanov:2004wn,Chekanov:2007tv,Chekanov:2008tn,Chekanov:2002yh,Andreev:2014zka} as well as the semi-exclusive production of dijet with high transverse energy $e p \to en j j X$~\cite{Breitweg:2000nk,Chekanov:2009ac,Aktas:2004gi}.

In our previous analysis~\cite{Shoeibi:2017lrl}, we have used the leading neutron productions data from ZEUS-02 \cite{Chekanov:2002pf} and H1 \cite{Aaron:2010ab} collaborations. 
The H1 semi-inclusive DIS structure function $F_2^{LN(3)} (\beta, Q^2; x_L)$ is measured within the DIS kinematic ranges of $6 < Q^2 < 100\,{\rm GeV}^2$, and  $1.5 \times 10^{-4} < x < 3 \times 10^{-2}$, and $0.02<y<0.6$ and neutron transverse
momentum $p_T < 0.2$ GeV.
The ratios of the measured $F_2^{LN(3)} (\beta, Q^2; x_L)$ to the inclusive $F_2^p(x, Q^2)$ presented in H1 measurement~\cite{Aaron:2010ab} show that these two structure functions have a similar $(x, Q^2)$ behavior. This result suggests the validity of hard scattering factorization presented in Eq.~\eqref{eq:factorization} and scale dependence of the nucleon FFs in Eq.~\eqref{eq:DGLAP}.

As we mentioned, our analysis is based on leading proton $e p \to e p X$ and leading neutron $e p \to e n X$ productions in semi-inclusive DIS at HERA. 
In Table~\ref{tab:tabledata}, we list all data sets included in {\tt STKJ17} global analysis along with the corresponding references, the kinematical coverage of $x_L$, $x$ and $Q^2$, the number of data points, and finally the normalization shifts ${\cal{N}}_n$ obtained in the QCD fit.

%
\begin{table*}[htb]
	\caption{List of all the leading neutron and leading proton productions data points above $Q^2 = 2.0 \, {\rm GeV}^2$ used in {\tt STKJ17} global analysis. For each
		dataset we provide the references, the kinematical coverage of $x_L$, $x$ and $Q^2$, the number of data points and the fitted normalization shifts ${\cal{N}}_n$ obtained in the fit.} \label{tab:tabledata}
	\begin{tabular}{l c c c c c c c}
Experiment & Observable & [$x_L^{\rm min}, x_L^{\rm max}$] & [$x^{\rm min}, x^{\rm max}$]  & $Q^2\,[{\rm GeV}^2]$  & \# of points & ${\cal N}_n$
\tabularnewline
\hline\hline
ZEUS-02~\cite{Chekanov:2002pf} & $F_2^{LN(3)}(x, Q^2; x_L)$ & [$0.24$--$0.92$]   & [$1.1 \times 10^{-5}$ -- $3.20 \times 10^{-3}$] & 7--1000 & \textbf{300} & 0.9974   \\		
ZEUS-06~\cite{Rinaldi:2006mf}  & $r^{LP(3)}(x, Q^2; x_L)$  &   [$0.575$--$0.890$] & [$8.5 \times 10^{-5}$ -- $8.29 \times 10^{-2}$] & 3.4--377 & \textbf{226} &  1.0012  \\		
ZEUS-09~\cite{Chekanov:2008tn} & $r^{LP(3)}(x, Q^2; x_L)$  &   [$0.370$--$0.895$] & [$9.6 \times 10^{-5}$ -- $3.25 \times 10^{-2}$] & 4.2--237 & \textbf{168} &  1.0004  \\		
\hline \hline
\multicolumn{1}{c}{\textbf{Total data}}  &  &  & &  &  \textbf{694}  \\  \hline
	\end{tabular}
\end{table*}
%
%

In next section, we briefly summarize the selection of data sets and the cuts imposed on them, and the treatment of experimental normalization uncertainties.

%
\subsubsection{ZEUS-02 data on leading neutron production}\label{sec:ZEUS-02Data}

As we summarized in Table~\ref{tab:tabledata}, we have used three data sets in our QCD fit. For the leading-neutron production in $e^+p$ collisions, we have used the data from ZEUS collaboration~\cite{Chekanov:2002pf}. At ZEUS experiment, a Forward Neutron Calorimeter (FNC) detected leading neutrons with the energy resolution of $\sigma/E = 65\%/\sqrt{E ({\rm GeV})}$, covering $0 \pm 0.8$ mrad. In order to reduce the systematic uncertainties, the neutron-tagged cross section, $ep \to e n X$, is measured relative to the inclusive DIS cross section, $ep \to eX$.
Leading neutron production has been studied for $x_L > 0.2$, production angle of the neutron $\theta_n < 0.8$, and the photon virtuality up to $Q^2 \sim 10^4 \, {\rm GeV}^2$. The transverse-momentum acceptance of the leading neutron in ZEUS experiment has been set to $p_T \approx x_L E_p \theta_n = 0.656 \, x_L  ({\rm GeV})$.

In order to minimize the systematics uncertainties, ZEUS collaboration at HERA studied the relationship between the leading neutron production and the inclusive $e^+ p$ scattering in terms of the ratio of the cross sections. The corresponding ratio $r^{LN(3)}$ in bins of width $\Delta x_L$ is then defined as~\cite{Chekanov:2002pf}

\begin{equation}\label{eq:rn}
r^{LN(3)} (x, Q^2; x_L) (1+ \Delta)  = \frac{F_2^{LN(3)}(x, Q^2; x_L)}{F_2^p(x, Q^2)} \Delta x_L \,,
\end{equation}

where the correction term $\Delta$ is small and can be neglected, and $F_2^p(x, Q^2)$ is the inclusive DIS structure function and neutron-tagged structure function $F_2^{LN(3)}(x, Q^2; x_L)$ is obtained by the integration over $p_T^2$ from Eq.~\eqref{intpT}. In our analysis, we have used the inclusive DIS structure function of {\tt GJR08} at NLO approximation~\cite{Gluck:2007ck}. Multiplying the $r^{LN(3)}$ ratios by a fit to the inclusive $F_2^p$, the ZEUS collaboration was also able to measured the neutron-tagged structure functions $F_2^{LN(3)}(x, Q^2; x_L)$ values for various bins of $x$, $x_L$ and $Q^2$.

In our global analysis, we have considered both leading neutron and leading proton production data. For ZEUS-02~\cite{Chekanov:2002pf} experiment, we have used the semi-inclusive cross section $\sigma_r^{LN(3)} (x, Q^2; x_L)$. Our theory setup also includes the negligible contributions from the longitudinal structure functions $F^{LN(3)}_2$.
Since in ZEUS-02 experiment, the cross section for the production of leading neutrons has been determined as a ratio $r^{LN(3)} (x, Q^2; x_L)$ relative
to the inclusive neutral current cross section, we have used Eq.~\eqref{eq:rn} to account this data.
The values of inclusive structure functions $F_2^p(x, Q^2)$ are obtained from the {\tt GJR08} parameterization of the parton density~\cite{Gluck:2007ck}.

\begin{figure}[htb]
	\begin{center}
		\vspace{0.5cm}
		\resizebox{0.80\textwidth}{!}{\includegraphics{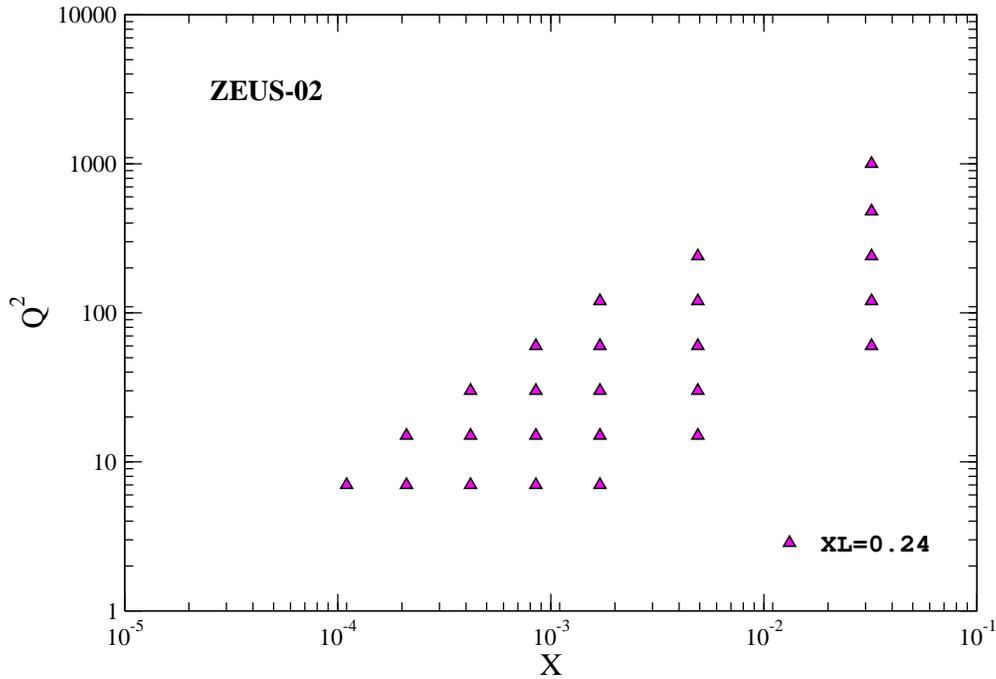}}   
		\caption{ (Color online) Nominal coverage of the ZEUS-02~\cite{Gluck:2007ck} data sets used in our global QCD fits for one selected bin of $x_L = 0.24$.}\label{fig:ZEUS-FLN}
	\end{center}
\end{figure}

In Fig.~\ref{fig:ZEUS-FLN}, we plot the nominal coverage of ZEUS-02 leading neutron data sets used in {\tt STKJ17} QCD fits for one representative bin of $x_L = 0.24$. 
The plot nicely summarizes the universal $x$, $x_L$, and $Q^2$ dependence of the leading neutron production at HERA.

%
\subsubsection{ZEUS-06 data on leading proton production}\label{sec:ZEUS-06Data}

In addition to the ZEUS-02 data on leading neutron production, we also include the ZEUS-06 data on leading proton production~\cite{Rinaldi:2006mf}.
Events of the type $e^+ p \to e^+pX$  with a final-state proton with $x_L > 0.6$ have been studied in $e^+p$ collisions at HERA using the ZEUS
detector~\cite{Chekanov:2002yh,Rinaldi:2006mf}, emphasizing the non-diffractive region. The high-energy leading protons with low transverse momentum carrying at least 60\% of the incoming-proton momentum and measured in the ZEUS leading proton spectrometer (LPS).
The ZEUS-06 data were taken during 1994-1995 using the HERA collider at DESY in which $E_e=27.5$ GeV positron collided with the $E_p=820$ GeV protons correspond to centre-of-mass energy of $\sqrt{s} = 301$ GeV. Data with different photon virtualities were used: $Q^2 < 0.02 \, {\rm GeV}^2$, $0.1 < Q^2 < 0.74 \, {\rm GeV}^2$ and $3 < Q^2 < 254 \, {\rm GeV}^2$ corresponding to the total integrated luminosity of ${\cal L} = 0.9, \, 1.85,$ and $3.38\,{\rm pb}^{-1}$.
The ratio of the cross section for leading proton production to the inclusive $e^+p$ cross section are given as a function of $x$, $x_L$, $Q^2$ and $p_T^2$.
The ratio $r^{LP(3)} (x, Q^2; x_L)$ is given by

\begin{equation}\label{eq:rp}
r^{LP(3)} (x, Q^2; x_L)  = \frac{F_2^{LP(3)}(x, Q^2; x_L)}{F_2^p(x, Q^2)}\,,
\end{equation}

\begin{figure}[htb]
	\begin{center}
		\vspace{0.5cm}
		\resizebox{0.80\textwidth}{!}{\includegraphics{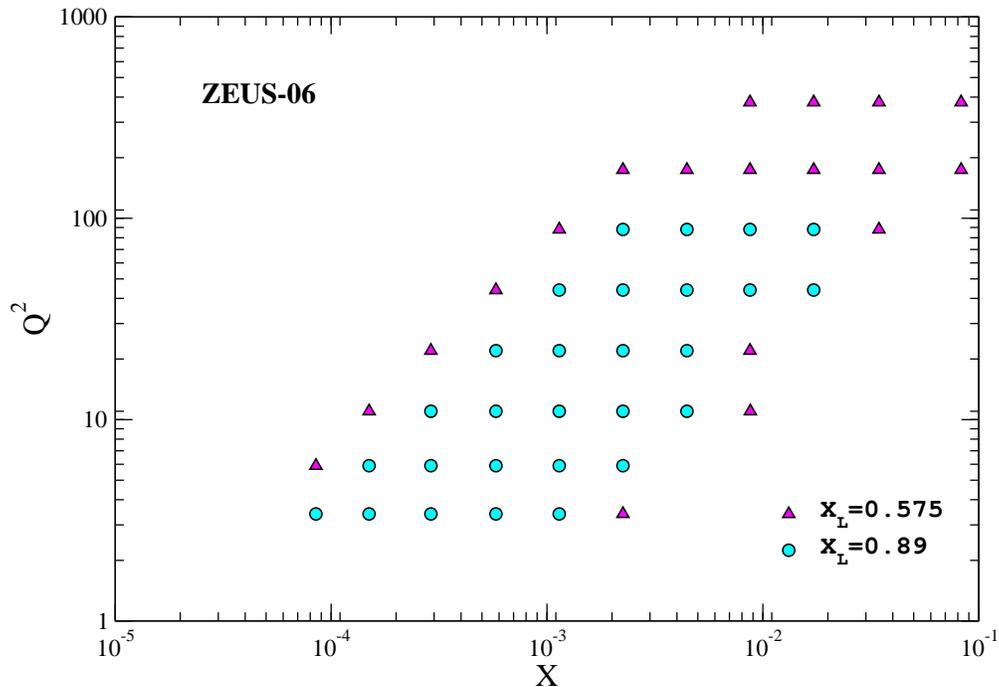}}   
		\caption{ (Color online) Nominal coverage of the ZEUS-06~\cite{Rinaldi:2006mf} data sets used in our global QCD fits for two selected bins of $x_L = 0.575$ and 0.89.}\label{fig:ZEUS-FLP}
	\end{center}
\end{figure}

The contributions from the longitudinal structure function $F_L$ is assumed to be the same for the inclusive DIS and the proton-tagged reactions.
For the inclusive DIS structure function, $F_2^p(x, Q^2)$, we have used the results of {\tt GJR08}~\cite{Gluck:2007ck}. 
Fig.~\ref{fig:ZEUS-FLP} shows the distribution of the leading proton production data in the $Q^2$ and $x$ plane for two representative bins of $x_L = 0.575$ and 0.89.  
We can see that the ZEUS-06 data set provides a rather wide coverage in the ($x$; Q$^2$) plane.

As we already mentioned, the ZEUS-06 leading proton production has been measured in several $x$ and $Q^2$ bins for $p^2_T < 0.50 \, {\rm GeV}^2$ which is different from ZEUS-02 leading neutron production which is measured in neutron transfer momentum of $p_T^2 < 0.43 \, x_L^2  ({\rm GeV})$. 
However the ZEUS-06 data cover the ZEUS-02 range in $x_L$, the distribution of the protons in ZEUS-06 is integrated up to $p^{\rm max}_T = 0.70 \, {\rm GeV}$.
Hence, for different values of $x_L$, the ZEUS-06 values must be reduced to account for the $p_T$ range measured by ZEUS-02.
For the leading neutron production at ZEUS experiment~\cite{Chekanov:2002pf}, distributions in $p_T$ are often parameterized by exponential $e^{-b(x_L) p_T^2}$ with a characteristic slope $b(x_L) = (16.3 x_L - 4.25) \, {\rm GeV}^{-2}$. For the leading proton production at ZEUS experiment~\cite{Chekanov:2002yh,Rinaldi:2006mf}, the values of the slope-parameters $b$ is independent of $Q^2$ and $x_L$ and mean value of $b$ is $b=6.6 \pm 0.6 \, ({\rm stat.}) \pm 0.8 \, ({\rm syst.}) \, {\rm GeV}^{-2}$ for $0.6 < x_L < 0.97$.

%
\subsubsection{ZEUS-09 data on leading proton production}\label{sec:ZEUS-09Data}

\begin{figure}[htb]
	\begin{center}
		\vspace{0.5cm}
		\resizebox{0.80\textwidth}{!}{\includegraphics{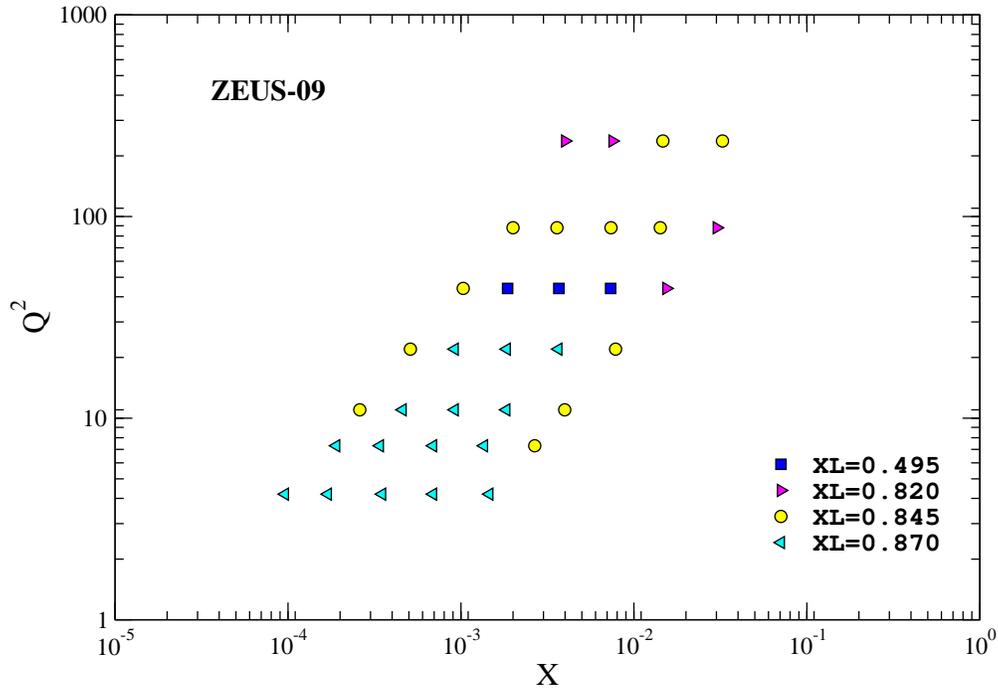}}   
		\caption{ (Color online) Typical kinematical coverage in the ($x$, Q$^2$) plane for the ZEUS-09 data sets included in our global QCD fits for four selected bins of $x_L = 0.495$, 0.820, 0.845 and 0.870.}\label{fig:ZEUS-09Data}
	\end{center}
\end{figure}

In addition to the mentioned data sets, we also have used the recent ZEUS data on leading proton production in semi-inclusive reaction $e^+p \to e^+p X$~\cite{Chekanov:2008tn}. 
The leading proton production cross section and its ratio to the inclusive DIS cross section  were studied with the ZEUS detector
at HERA with an integrated luminosity of 12.8 $pb^{-1}$. 
In this experiment, leading protons carried a large fraction of the incoming 
proton energy, $x_L > 0.32$, and its transverse momentum squared satisfied $p_T^2 < 0.5 \, {\rm GeV}^2$. An approximately 24\% of DIS events in this measurement have a leading proton.
In Fig.~\ref{fig:ZEUS-09Data} we show the kinematic coverage in the ($x$; Q$^2$) plane of the ZEUS-09 datasets included in our global analysis. 
The plot nicely summarizes the universal $x$, $x_L$, and $Q^2$ dependence of the leading proton productions using ZEUS detector at HERA collider.

%
\subsection{$\chi^2$ analysis and uncertainties of nucleon FFs}\label{uncertainties}

The total $\chi^2$ is calculated in comparison with the leading nucleon data for the nucleon FFs in Eq.~\eqref{eq:PDFQ0-New}.
The theoretical functions should be obtained at the same experimental $x$, $x_L$ and $Q^2$ points for calculating $\chi^2$.
As we explained in Sec.~\ref{sec:Evolution}, the $Q^2$ evolution is calculated by the DGLAP evolution equations of Eq.~\eqref{eq:DGLAP}.
The simplest method to calculate the total $\chi^2 (\{\zeta_i\})$ for independent sets of unknown fit parameters $\{\zeta_i\}$ is as follow

\begin{eqnarray}\label{eq:chi2-1}
\chi^2 (\{\zeta_i\}) = \sum_{i}^{ n^{data} } \frac{ (D^{data}_i - T^{theory}_i (\{\zeta_i\}))^2 } { (\sigma^{data}_i)^2 } \,,
\end{eqnarray}

where $D^{data}_i$ and $T^{theory}_i$ are experimental and theoretical values of leading nucleon observables, respectively, at the same experimental $x$, $x_L$ and $Q^2$ points. The experimental errors are calculated from systematic and statistical errors added in quadrature, $\sigma^{data}_i = \sqrt{(\sigma^{sys}_i)^2 + (\sigma^{stat}_i)^2}$. The unknown parameters are determined so as to
obtain the minimum $\chi^2$. The optimization of the functions is done by the CERN program MINUIT~\cite{James:1994vla}.

Since most experiments come with additional information on the fully correlated normalization uncertainty $\Delta {\cal N}_n$, the simple $\chi^2$ definition in Eq.~\eqref{eq:chi2-1} need to be modified in order to account for such normalization uncertainties. In this case and in order to determine the best fit parameters of Eq.~\eqref{eq:PDFQ0-New}, we need to minimize the $\chi^2_{\rm global}(\{\zeta_i\})$ function with the free unknown parameters. This function is given by,
\begin{equation}\label{eq:chi2}
\chi_{\rm global}^2 (\{\zeta_i\}) = \sum_{n=1}^{n^{exp}} w_n  \chi_n^2\,,
\end{equation}
where $w_n$ is a weight factor for the $n^{th}$ experiment and
\begin{eqnarray}\label{eq:chi2global}
\chi_n^2 (\{\zeta_i\})=\left( \frac{1 -{\cal N}_n }{\Delta{\cal N}_n}\right)^2+\sum_{j=1}^{N_n^{data}} \left(\frac{ ( {\cal N}_n  \, {D}_j^{data} - {T}_j^{theory}(\{\zeta_i\})}{{\cal N}_n \, \delta {D}_j^{data}} \right)^2\,,
\end{eqnarray}
where $n^{\rm exp}$ correspond to the individual experimental data sets and $N^{\rm data}_n$ correspond to the number of data points in each data set.
The normalization factors $\Delta {\cal N}_n$ in Eq.~\eqref{eq:chi2global} can be fitted along with the fitted parameters $(\{\zeta_i\})$ of Eq.~\eqref{eq:PDFQ0-New} and then keep fixed.
The obtained normalization factors $\Delta {\cal N}_n$ are presented in Table.~\ref{tab:tabledata}.
In order to illustrate the effects arising from the use of the different data sets, in Tables.~\ref{tab:chisquaredZEUS-02}, ~\ref{tab:chisquaredZEUS-06} and ~\ref{tab:chisquaredZEUS-09} we show the $\chi/n^{data}$ for each bin of $x_L$.
These tables illustrate the quality of our QCD fits to leading nucleon production data at NLO accuracy in terms
of the individual $\chi^2$-values obtained for each experiment. The total $\chi^2/N_{pts}$ for the resulting fit to the ZEUS-02, ZEUS-06 and ZEUS-09 datasets is $770.679/677 = 1.138$.

%
\begin{table}[htbp]
	\centering
	{\footnotesize
		\begin{tabular}{c | c | c | c c}
			\hline \hline
			Experiment      & $x_L$    & $\chi^2$    & $n^{data}$  \\ \hline  
			& $x_L$ = 0.24  & 15.808      &   25        &   \\
			& $x_L$ = 0.31  & 17.862      &   25        &   \\
			& $x_L$ = 0.37  & 21.009      &   25        &   \\
			& $x_L$ = 0.43  & 22.463      &   25        &   \\
    		& $x_L$ = 0.49  & 24.821      &   25        &   \\
ZEUS-02~\cite{Chekanov:2002pf}     & $x_L$ = 0.55  & 17.551      &   25        &    \\
			& $x_L$ = 0.61  & 23.509      &   25        &    \\
			& $x_L$ = 0.67  & 28.305      &   25        &    \\
			& $x_L$ = 0.73  & 62.006      &   25        &    \\
			& $x_L$ = 0.79  & 55.432      &   25        &    \\
			& $x_L$ = 0.85  & 58.254      &   25        &    \\
			& $x_L$ = 0.92  & 41.693      &   25        &    \\  \hline
			All data sets & & \textbf{ 388.713  } & \textbf{ 300 }  \\    \hline  
			\hline
		\end{tabular}
	}
	\caption{The values of $\chi^2/n^{data}$ for the ZEUS-02 dataset~\cite{Chekanov:2002pf} included in the {\tt STKJ17} global QCD analysis. More detailed discussion of the description of the individual data sets, and the definitions of $\chi^2 (\{\xi_i\})$ are contained in the text.}
	\label{tab:chisquaredZEUS-02}
\end{table}

%
\begin{table}[htbp]
	\centering
	{\footnotesize
		\begin{tabular}{c | c | c | c c}
			\hline \hline
			Experiment &   $x_L$        & $\chi^2$     & $n^{data}$   \\ \hline    
			& $x_L$ = 0.575 & 12.076       &    19        &     \\
			& $x_L$ = 0.59 & 15.980        &    28        &     \\
			& $x_L$ = 0.65 & 24.398        &    29        &     \\
ZEUS-06~\cite{Rinaldi:2006mf}		& $x_L$ = 0.71 & 21.284        &    28        &     \\
			& $x_L$ = 0.725 & 35.788       &    19        &     \\
			& $x_L$ = 0.77 & 26.084        &    28        &     \\
            & $x_L$ = 0.83 & 74.326        &    28        &     \\
			& $x_L$ = 0.875 & 30.024       &    19        &     \\
			& $x_L$ = 0.89 & 53.001        &    28        &     \\    \hline
			All data sets & & \textbf{ 292.961  } & \textbf{ 226 }   \\ \hline  
			\hline
		\end{tabular}
	}
	\caption{The values of $\chi^2/n^{data}$ for the ZEUS-06 dataset~\cite{Rinaldi:2006mf} included in the {\tt STKJ17} global QCD analysis], see details in Table.~\ref{tab:chisquaredZEUS-02}.}
	\label{tab:chisquaredZEUS-06}
\end{table}

%
\begin{table}[htbp]
	\centering
	{\footnotesize
		\begin{tabular}{c | c | c | c c}
			\hline \hline
			Experiment &  $x_L$   & $\chi^2$  & $n^{data}$   \\ \hline   
			& $x_L$ = 0.37 & 7.399       &    15        &    \\
			& $x_L$ = 0.392 & 1.62       &    3         &   \\
			& $x_L$ = 0.395 & 4.763       &    12        &    \\
			& $x_L$ = 0.42 & 2.352       &    4        &    \\
    	    & $x_L$ = 0.47 & 19.331        &    15      &       \\
			& $x_L$ = 0.495 & 2.674       &    3        &    \\
			& $x_L$ = 0.545 & 6.022       &    12        &    \\
			& $x_L$ = 0.57 & 4.569       &    15       &    \\
			& $x_L$ = 0.595 & 2.197       &    3         &   \\
ZEUS-09~\cite{Chekanov:2008tn}  	& $x_L$ = 0.62 & 2.359       &    4         &   \\
			& $x_L$ = 0.67 & 4.964       &    15         &   \\
			& $x_L$ = 0.695 & 8.826       &    15        &    \\
			& $x_L$ = 0.77 & 6.82       &    15        &    \\
			& $x_L$ = 0.795 & 1.238       &    3        &    \\
			& $x_L$ = 0.82 & 1.13       &    4         &   \\
			& $x_L$ = 0.845 & 5.275       &    12         &   \\
			& $x_L$ = 0.87 & 2.216       &    15         &   \\
			& $x_L$ = 0.895 & 5.064       &    3        &    \\ \hline
			All data sets & & \textbf{ 89.005  } & \textbf{ 168 }    \\    
			\hline \hline 
		\end{tabular}
	}
	\caption{The values of $\chi^2/n^{data}$ for the ZEUS-09 dataset~\cite{Chekanov:2008tn} included in the {\tt STKJ17} global QCD analysis, see details in Table.~\ref{tab:chisquaredZEUS-02}.}
	\label{tab:chisquaredZEUS-09}
\end{table}

The determination of PDFs, polarized PDFs, nuclear PDFs as well as nucleon FFs through a QCD fit to the experimental data is a procedure that necessarily implies a variety of assumptions, mostly concerning their input parameterization and the propagation
of the experimental uncertainties into them~\cite{Gao:2017yyd}.
In recent years, the assessment of uncertainties has seen significant progress in the QCD analyses.
Among the different approaches the Hessian method, the Lagrange multiplier technique (LM) and Neural Network (NN) are the
most reliable ones.
The ``Hessian method'' in which use have used to extract the uncertainties of the neutron FFs as well as the corresponding observables were estimated in Refs.~\cite{deFlorian:2011fp,Hou:2016sho,MoosaviNejad:2016ebo,Khanpour:2016uxh,Khanpour:2016pph,Shoeibi:2017lrl,Martin:2002aw,Martin:2009iq,Pumplin:2001ct}. Although technical details of the Hessian method are described in these references, outline of this method is
explained here because it is used in our analysis.

As we discussed, our method is the $\chi_n^2 (\{\zeta_i\})$ fitting procedure used in the global QCD analysis and for the determination of the uncertainties, we have used the well-known ``Hessian'' or error matrix approach.
This method confirms that the {\tt STKJ17} fitting methodology used in this QCD analysis can faithfully reproduce the input nucleon FFs in the region where the leading nucleon production data are sufficiently constraining.
The fit parameters are denoted $\zeta_i$ ($i$=1, 2, ..., $N$), where $N$ is the total number of the fitted parameters. One can expand $\chi^2$ around the minimum $\chi^2$ point $\hat \zeta$ as

\begin{equation}
\Delta \chi^2 (\zeta) = \chi^2(\hat{\zeta} + \delta \zeta)-\chi^2(\hat{\zeta}) = \sum_{i,j} H_{ij}\delta \zeta_i \delta \zeta_j \ ,
\end{equation}

where $H_{ij}$ are the elements of the Hessian matrix which is given by

\begin{equation}
H_{ij} = \frac{1}{2} \frac{\partial^2 \chi^2}{\partial \zeta_i \partial \zeta_j} \mid_{min}  \,.
\end{equation}

The confidence region normally is given in the parameter space by supplying a value of $\Delta \chi^2$.
In the standard parameter-fitting criterion, the errors are given by the choice of the tolerance $T=\Delta \chi^2 =1$.
It is also known that the confidence level (C.L.) is 68\% for $\Delta \chi^2 =1$ if the number of the fitted parameters is one $N=1$. It is important to know that for the general case in which the number of fitted parameters is $N > 1$, the $\Delta \chi^2$ value needs to be calculated to determine the size of the uncertainties.
This indicates that our fitting methodology as well as the uncertainties determination correctly propagate the experimental uncertainty of the
data into the uncertainties of the fitted nucleon FFs.

The determination of the size of uncertainties have been done by applying the Hessian
method based on the correspondence between the confidence level $P$ and $\chi^2$ with the number of fitting parameters $N$.
The confidence level $P$ is given by,

\begin{equation}
P = \int_0^{\Delta \chi^2}\frac{1}{2\,\Gamma(N/2)} \left(\frac{\chi^2}{2}\right)^{\frac{N}{2}-1} e^{\left(-\frac{\chi^2}{2} \right)} d\,\chi^2 \,,
\end{equation}

where $\Gamma$ is the Gamma function. The value of $\Delta \chi^2$ in above equation is taken so that the confidence level (C.L.) becomes the one-$\sigma$-error range, namely $P=0.68$. Similarly, for the 90th percentile, we have $P=0.90$. The value of $\Delta \chi^2$ is then numerically calculated by using equation above.

The Hessian matrix or error matrix can be obtained by running the subroutine the CERN program library MINUIT~\cite{James:1994vla}.
The uncertainty on any observables ${\cal O}$, which is an attributive function of the input parameters obtained in
the QCD analysis at the input scale Q$_0^2$, is obtained by applying the Hessian method.
Having at hand the value of $\Delta \chi^2$, and derivatives of the observables with respect to the fitted parameters, the Hessian method gives the uncertainties on any observables ${\cal O}$ as,

\begin{equation}
[\Delta {\cal O}_i]^2=\Delta \chi^2 \sum_{j,k}	\left( \frac{\partial {\cal O}_i (\zeta)}{\partial \zeta_j}  \right)_{\hat \zeta}
C_{j,k}	\left( \frac{\partial {\cal O}_i (\zeta)}{\partial \zeta_k}  \right)_{\hat \zeta} \,,
\end{equation}

where $C_{j,k}$ is the inverse of the Hessian matrix, $H_{jk}^{-1}$.
For estimation of uncertainties at an arbitrary $Q^2$, the obtained gradient terms are evolved by the DGLAP evolution kernel,
and then the nucleon FFs uncertainties as well as the uncertainties of other observables, such as leading nucleon structure function or cross sections, are calculated.
We should notice here that a set of uncertainties due to the theoretical method have been included in our analysis. Specially
we considered 5\% uncertainty due to {\tt GJR08} PDFs in which we have used in our definition for the leading nucleon FFs at the input scale, $Q_0^2 = 2 \, {\rm GeV}^2$.

In the next section, we present the main results of this work, namely the ``{\tt STKJ17}'' set of nucleon FFs at NLO approximation. First we
discuss the resulting nucleon FFs and their uncertainties. Then, we show the quality of the fits and compare the {\tt STKJ17} predictions to the fitted leading nucleon production data sets.

%
\section{The results of global QCD analysis}\label{sec:results}

In this section we will present and discuss in depth the main results of {\tt STKJ17} QCD global analysis of nucleon FFs. First,
we present the optimum fit parameters and the constraints applied to control the nucleon FFs parameters.
Next, the newly obtained nucleon FFs and their uncertainty estimates are shown. The quality of the fit to ZEUS-02, ZEUS-06 and ZEUS-09 datasets and potential open issues and tensions
among the different sets of data are illustrated and discussed in this section.

{\tt STKJ17} fitted parameters in the NLO approximation at the input scale $Q_0^2 = 2 \, {\rm GeV}^2$ obtained from the best fit to the combined ZEUS-02 leading neutron data, and ZEUS-06 and ZEUS-09 leading proton datasets are listed in Table.~\ref{fit-parameters}. All these datasets could provide sensitivity to the flavor
separation of the nucleon FFs that was not available in the our previous analysis~\cite{Shoeibi:2017lrl}.

\begin{table*}[htbp]
	\caption{ Parameter values $\{\zeta_i\}$ for {\tt STKJ17} QCD analysis at the input scale $Q_0^2 = 2  \, {\rm GeV}^2$ obtained from QCD fit to the ZEUS-02, ZEUS-06 and ZEUS-09 datasets. The details of the $\chi^2 (\{\zeta_i\})$ analysis and
		the constraints applied to control the nucleon FFs parameters are contained in the text. \label{fit-parameters}}
	\begin{tabular}{l|ccccccccc}
		\hline  \hline
		Parameters                           & ${\cal N}$          & $A$                 & $B$                    &  $C$          &  $D$   \\   \hline  \hline
		${\cal W}_{u_v}(x_L)$                & $14.995 \pm 0.272$  & $3.866 \pm 0.121$   & $1.700 \pm 0.037$      &  $0.0$        & $0.0$  \\
		${\cal W}_{d_v}(x_L)$                & $11.004 \pm 0.113$  & $3.866 \pm 0.121$   & $1.50^*$               &  $0.0$        & $0.0$  \\
		${\cal W}_{\Delta}(x_L)$             & $47.026 \pm 9.734$  & $3.255 \pm 0.301$   & $0.948 \pm 0.107$      &  $0.0$        & $0.0$  \\
${\cal W}_{s}(x_L)={\cal W}_{\bar s}(x_L)$ & $0.308 \pm 0.025$   & $0.708 \pm 0.075$   & $1.544 \pm 0.049$      &  $17.387 \pm 1.988$   & $7.063 \pm 0.314$  \\
		${\cal W}_g(x_L)$                    & $1.750 \pm 0.269$   & $2.379 \pm 0.179$   & $2.426 \pm 0.101$      &  $38.074 \pm 8.965$   & $14.850 \pm 1.321$ \\  \hline
		$\alpha_S(Q_0^2)$                    &                     &                     & $0.356^*$              &               &  \\ 
		$\alpha_S(M_Z^2)$                    &                     &                     & $0.118$                &               &  \\ 
		$\chi^2/N_{pts}$                     &                     &                     & $770.679/677 = 1.138$  &               &  \\  \hline   \hline
	\end{tabular}
\end{table*}

In this analysis, it was difficult to determine all unknown parameters of the valence quark densities, the anti-quark, the strange sea and gluon functions of Eq.~\eqref{eq:PDFQ0-New}, so that we decided to fix some parameters. It indicates that the data are not sensitive to all parton species at this stage even in the NLO analysis. For valence quark and ${\cal W}_{\Delta}(x_L)$ densities, we prefer to set $C_i$ and $D_i$ to zero.  
The values without errors in Table.~\ref{fit-parameters} have been fixed after the first minimization since the data do not constrain these unknown parameters well enough.
With the available leading nucleon data sets, for example, the fit can not constrain distinct strange-quark fracture functions. We therefore assume a symmetric strange-quark distributions. As we discussed in section.~\ref{sec:Input}, we considered ${\cal W}_{s}(x_L) = {\cal W}_{\bar s}(x_L) $ = $\frac{1}{4} \times {\cal W}_{(\bar{d}+\bar{u})}(x_L)$. 

\begin{figure}[htb]
	\begin{center}
		\vspace{0.5cm}
		\resizebox{0.48\textwidth}{!}{\includegraphics{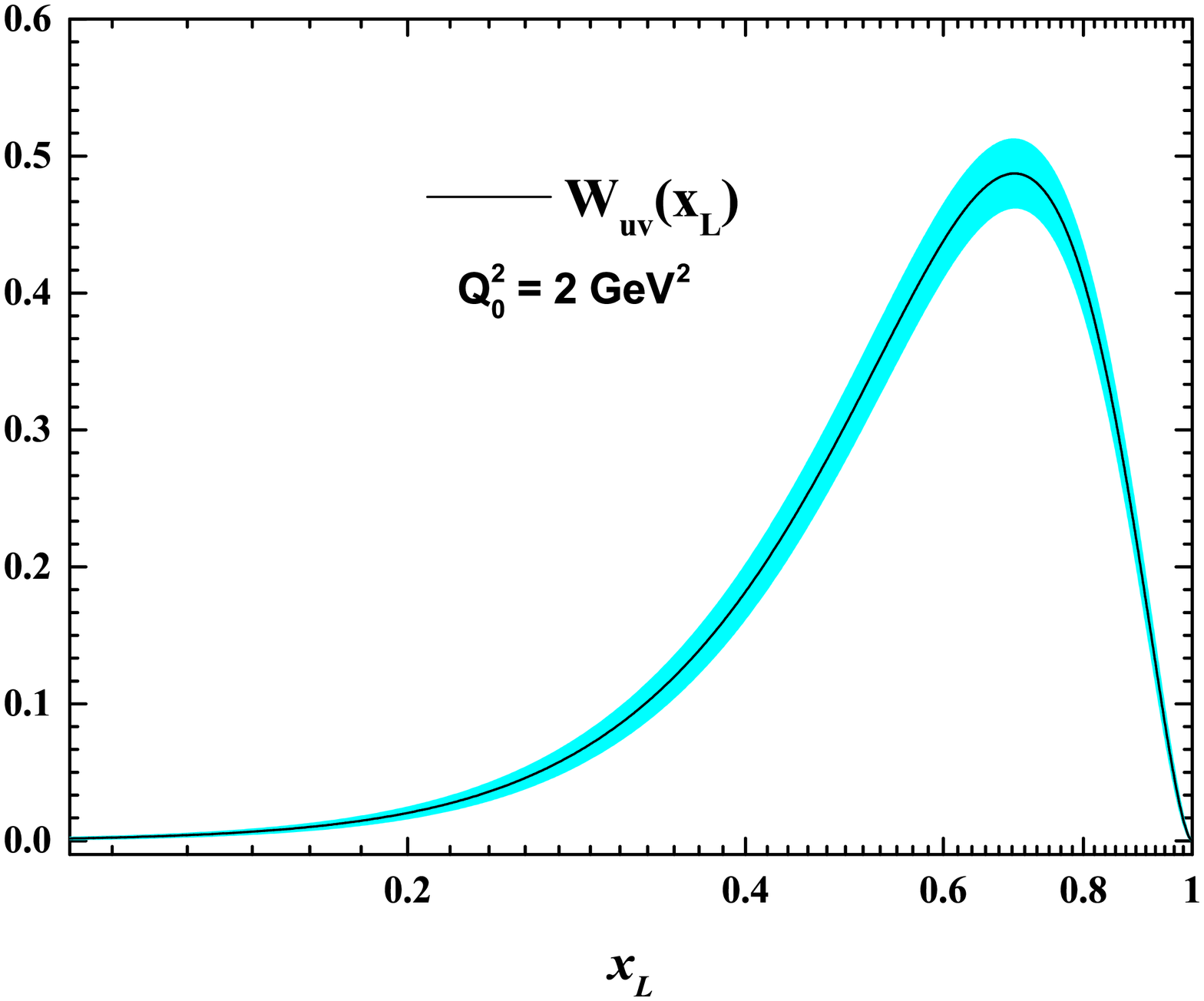}}    
		\resizebox{0.48\textwidth}{!}{\includegraphics{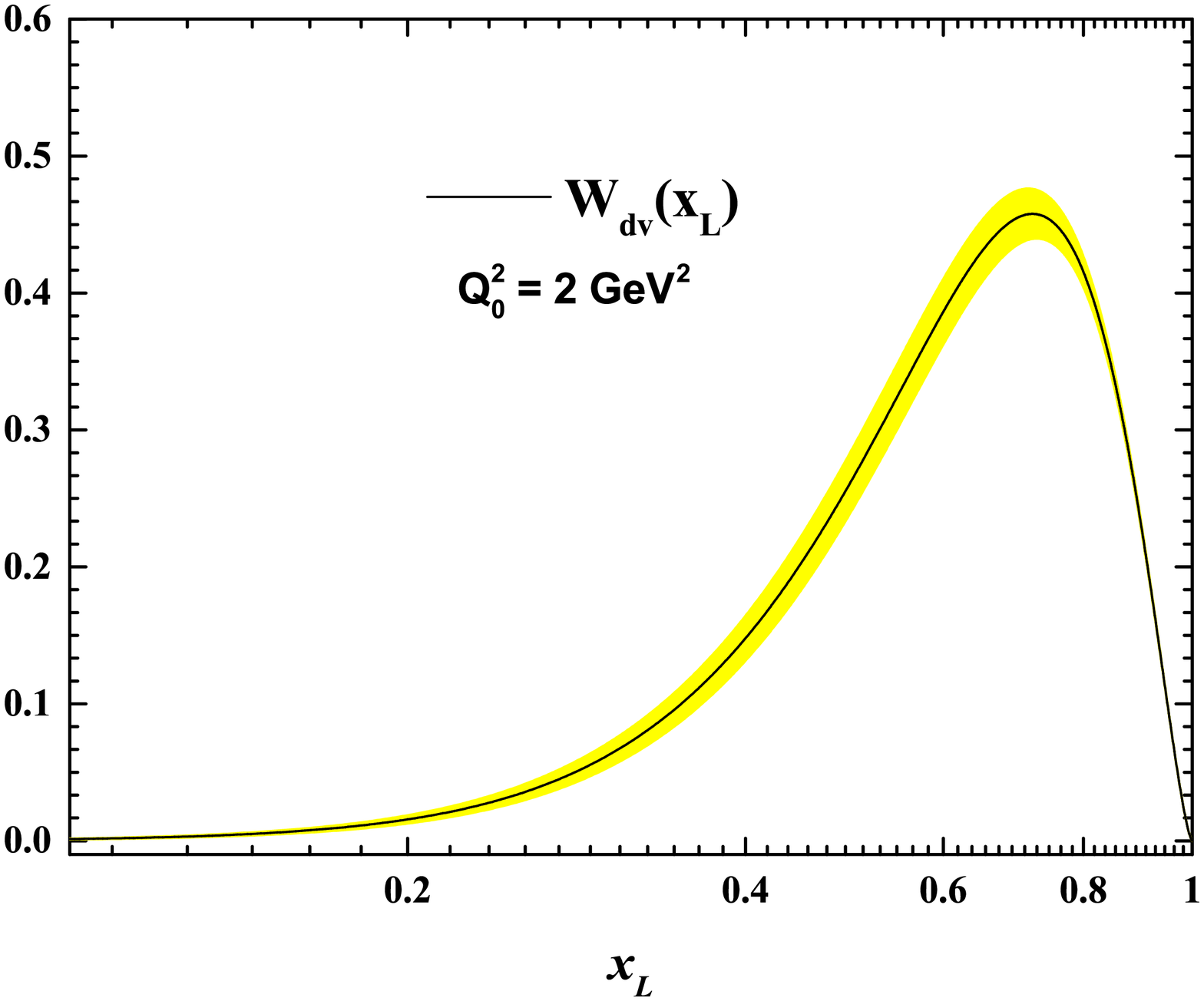}}   
		\caption{ (Color online) Our results for the ${\cal W}_{u_v}(x_L)$ and ${\cal W}_{d_v}(x_L)$ as a function of $x_L$ at the input scale $Q_0^2 = 2 \, {\rm GeV}^2$. }\label{fig:weight}
	\end{center}
\end{figure}

In Fig.~\ref{fig:weight} we present our results for the ${\cal W}_{u_v}(x_L)$ and ${\cal W}_{d_v}(x_L)$ as a function of $x_L$ at the input scale $Q_0^2 = 2 \, {\rm GeV}^2$.
The shaded bands correspond to the uncertainty estimates at 68\% confidence level (C.L.) for $\Delta \chi^2 = 1$. As can be seen from the plot, both ${\cal W}_{u_v}(x_L)$ and ${\cal W}_{d_v}(x_L)$ have similar pattern and pick at $x_L \approx 0.65$. For the leading nucleon production at HERA, the shape of the $x_L$ spectra is independent
of the kinematic variables $x$ and Q$^2$ confirming the hypothesis of limiting fragmentation~\cite{Benecke:1969sh,Chou:1994dh}.

\begin{figure*}[htb]
	\begin{center}
		\vspace{0.5cm}
		\resizebox{0.80\textwidth}{!}{\includegraphics{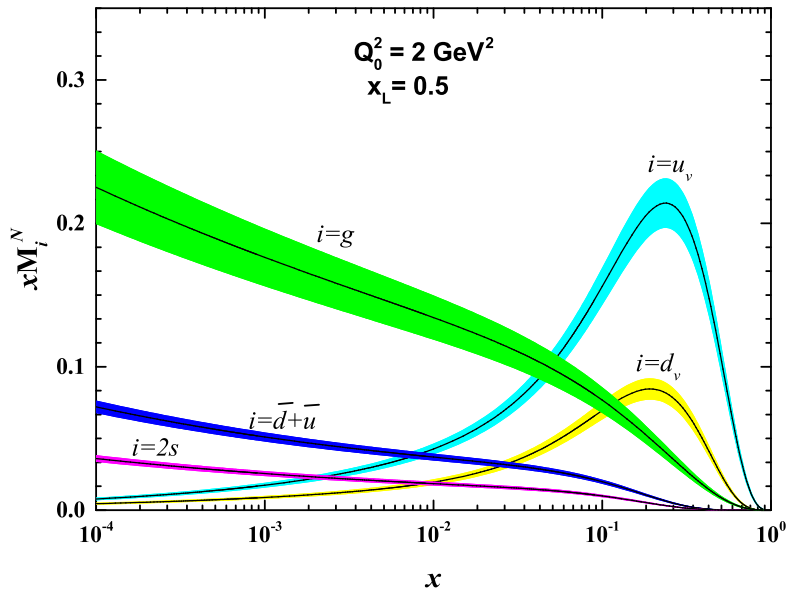}}   
		\caption{ (Color online) The nucleon FFs $x {\cal M}_{i} (x, Q^2; x_L)$ for all parton species resulting from our QCD analysis at the input scale, which is taken to be $Q_0^2 = 2 \, {\rm GeV}^2$. The results have been presented for a fixed value of $x_L = 0.5$. }\label{fig:PDF-XL=05}
	\end{center}
\end{figure*}

\begin{figure*}[htb]
	\begin{center}
		\vspace{0.5cm}
		\resizebox{0.80\textwidth}{!}{\includegraphics{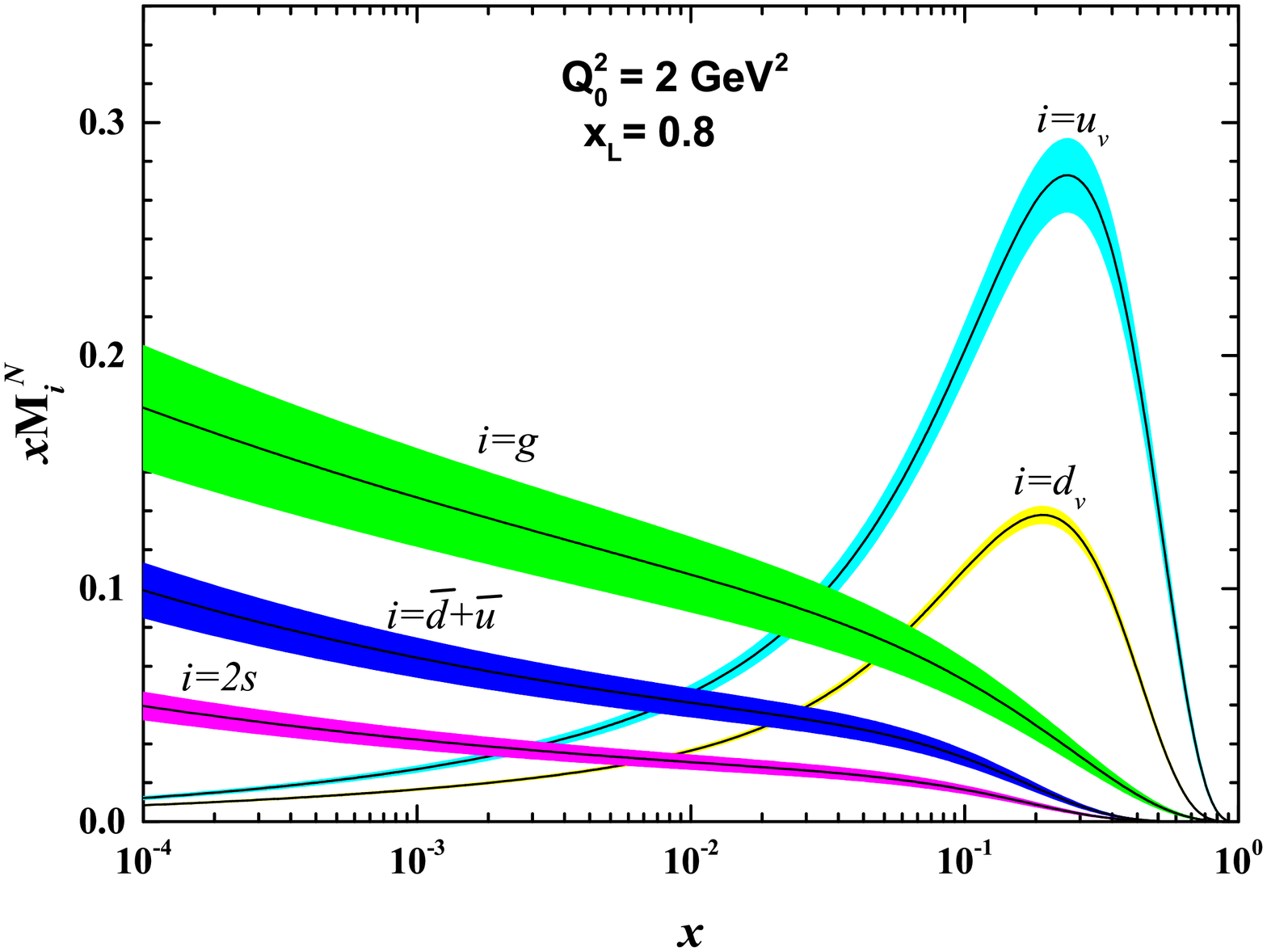}}   
		\caption{ (Color online) The nucleon FFs $x {\cal M}_{i} (x, Q^2; x_L)$ for all parton species resulting from our QCD analysis at the input scale, which is taken to be $Q_0^2 = 2 \, {\rm GeV}^2$. The results have been presented for a fixed value of $x_L = 0.8$. }\label{fig:PDF-XL=08}
	\end{center}
\end{figure*}

The nucleon FFs $x {\cal M}_{i} (x, Q^2; x_L)$ for all parton species resulting from our QCD analysis are shown in Fig.~\ref{fig:PDF-XL=05} and \ref{fig:PDF-XL=08} at the input scale, which is taken to be $Q_0^2 = 2 \, {\rm GeV}^2$. The shaded bands correspond to the uncertainty estimates at 68\% confidence level (C.L.).
In terms of uncertainties, the gluon FFs is less well constrained by the leading nucleon production data. As can be inferred from the figures, the error bands for light quark FFs at $x_L=0.8$ are bigger than those of $x_L=0.5$.


\begin{figure}[htb]
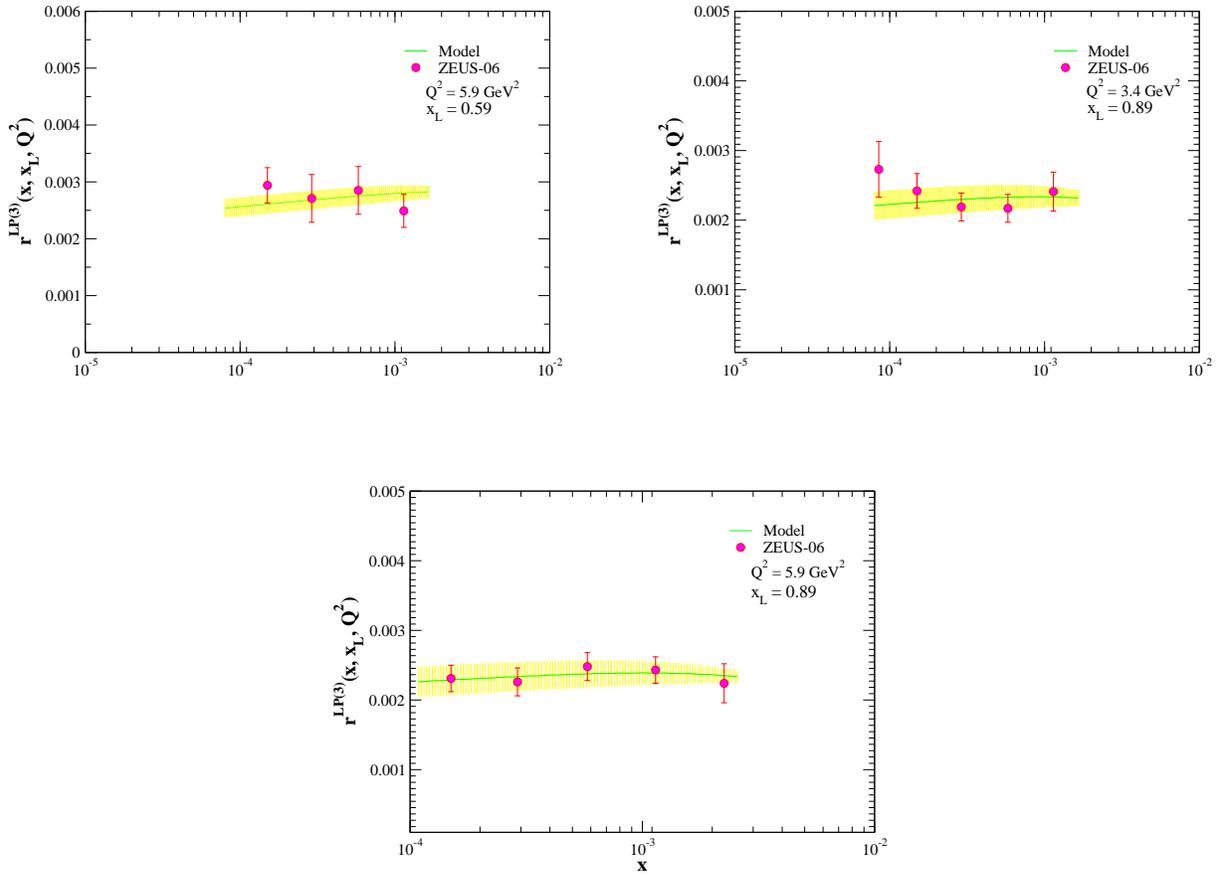

	\begin{center}
		\vspace{0.5cm}
		\resizebox{0.45\textwidth}{!}{\includegraphics{Ratio-xL059-Q2=5-9.eps}}  \vspace{1.2cm} \hspace{1.cm}  
		\resizebox{0.45\textwidth}{!}{\includegraphics{Ratio-xL089-Q2=3-4.eps}}     
		\resizebox{0.45\textwidth}{!}{\includegraphics{Ratio-xL089-Q2=5-9.eps}}   
		\caption{ (Color online) Our theory predictions for the structure function ratio $r^{LP(3)} (x, Q^2; x_L)  = \frac{F_2^{LP(3)}(x, Q^2; x_L)}{F_2^p(x, Q^2)}$ and its uncertainties at 68\% C.L. in comparison with the ZEUS-06 data~\cite{Rinaldi:2006mf}. }\label{fig:Ratio}
	\end{center}
\end{figure}

In Fig.~\ref{fig:Ratio} we present a detailed comparison of our theory prediction for the  $r^{LP(3)} (x, Q^2; x_L)  = \frac{F_2^{LP(3)}(x, Q^2; x_L)} {F_2^p(x, Q^2)}$ and its uncertainties at 68\% C.L. with the ZEUS-06 data already included.
The ratio is estimated by dividing the tagged-proton structure function $F^{LP(3)}_2$ computed with our nucleon FFs obtained in the present analysis by the inclusive DIS structure function $F_2^p$ extracted from the {\tt GJR08} PDFs.
In general, the agreement of the fit with this data is excellent in the $x$, $x_L$ and Q$^2$ covered by the ZEUS experiments.

As we mentioned, the hypothesis of limiting fragmentation predicts that the lepton variables $x$ and Q$^2$ completely separate from the baryon variables $x_L$.
The observation that $r^{LP(3)} (x, Q^2; x_L)$ is approximately constant over a large kinematic range in the
leptonic variables is in good agreement with ZEUS-06 data~\cite{Rinaldi:2006mf}.  Our results show that, for nucleon in DIS, hypothesis of limiting fragmentation works well at medium and high values of $x_L$.

\begin{figure}[htb]
	\begin{center}
		\vspace{0.5cm}
		\resizebox{0.45\textwidth}{!}{\includegraphics{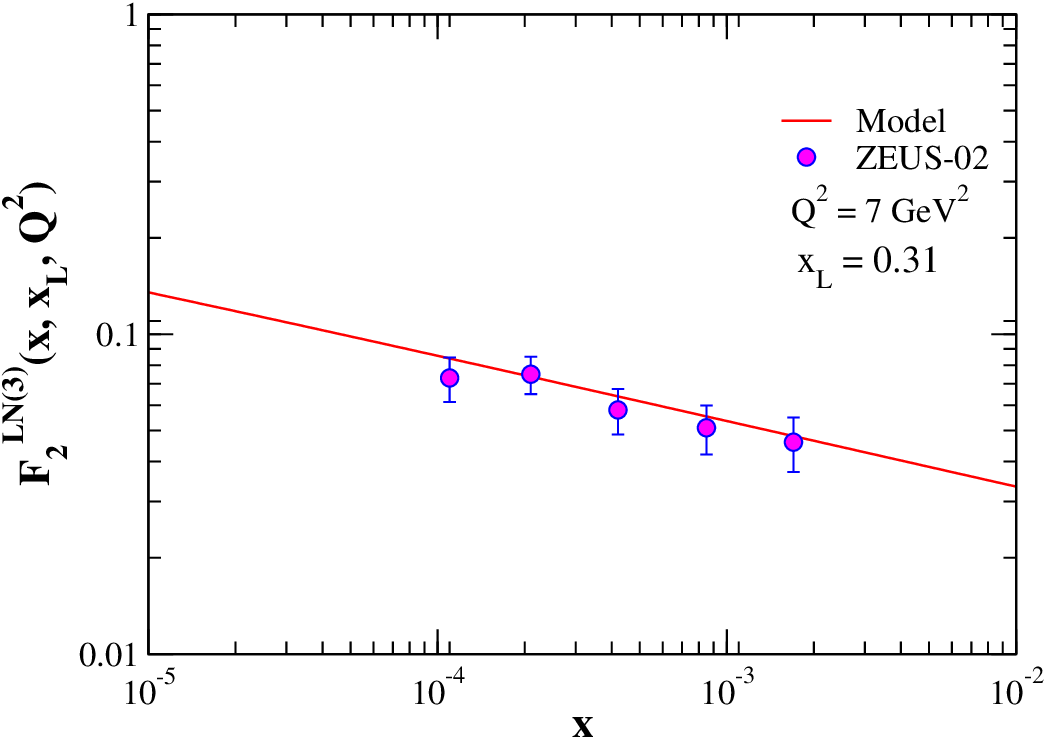}}    \vspace{1.cm} 
		\resizebox{0.45\textwidth}{!}{\includegraphics{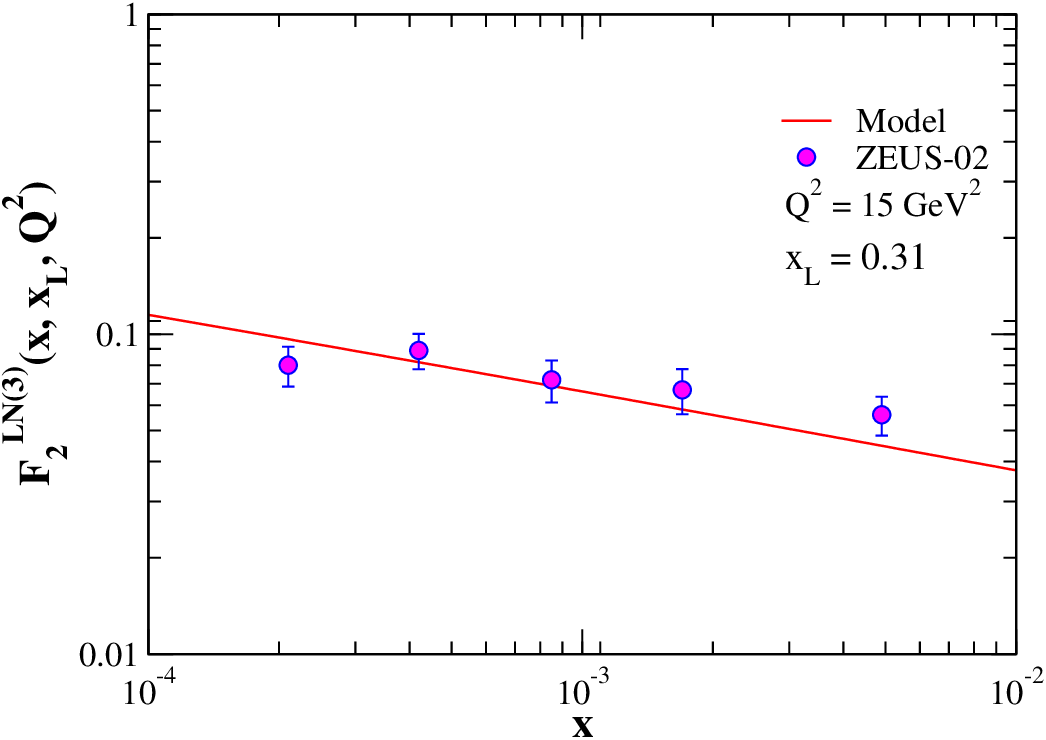}}    
		\resizebox{0.45\textwidth}{!}{\includegraphics{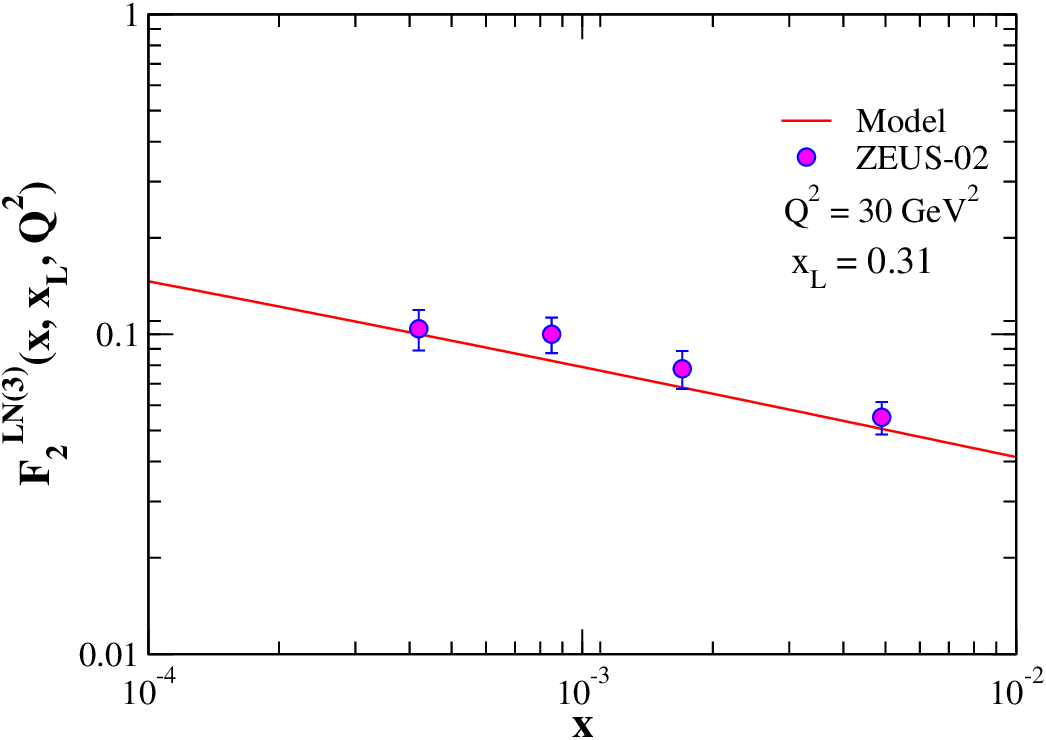}}   \vspace{1.cm} 
		\resizebox{0.45\textwidth}{!}{\includegraphics{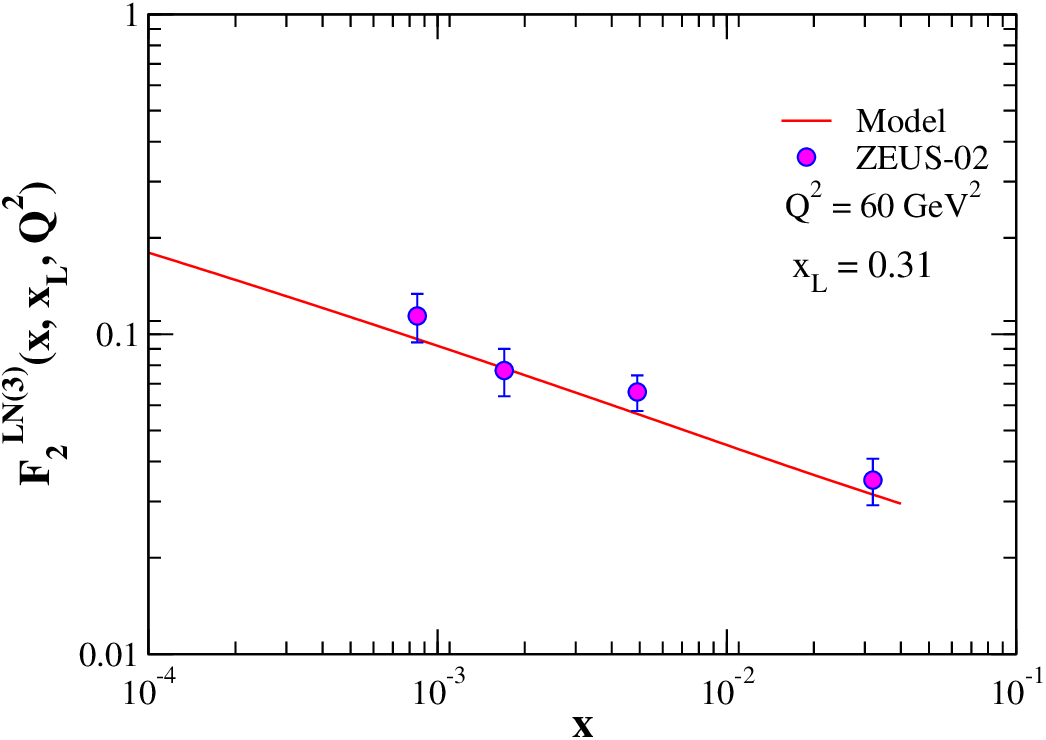}}   
		\resizebox{0.45\textwidth}{!}{\includegraphics{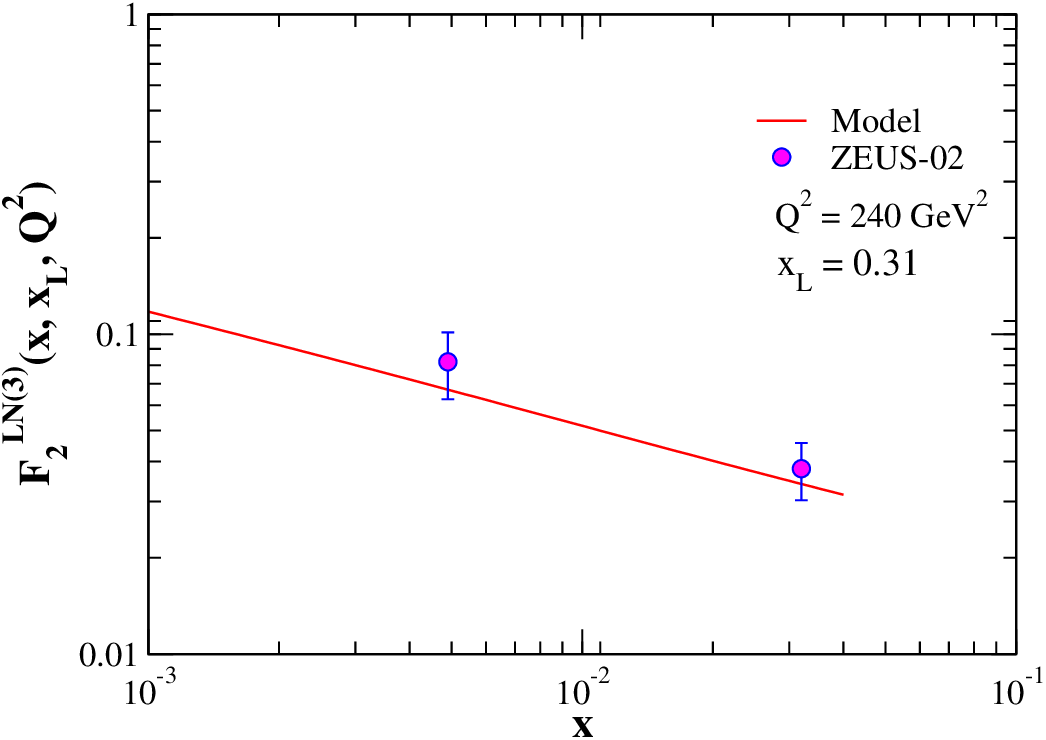}}   
		\caption{ (Color online) The tagged-neutron structure function $F^{LN(3)}_2$ as a function of $x$ for some selected values of Q$^2$ at fixed value of $x_L = 0.31$. Our theory predictions have been compared with the ZEUS-02 leading neutron data~\cite{Chekanov:2002pf}. }\label{fig:Comparison1}
	\end{center}
\end{figure}

Fig.~\ref{fig:Comparison1} shows our theory perditions for the tagged-neutron structure function $F^{LN(3)}_2$ as a function of $x$ for some selected values of Q$^2$ at fixed value of $x_L = 0.31$. The results are correspond to five different values of $Q^2$ = 7, 15, 30, 60 and 240 GeV$^2$. Our theory predictions have been compared with the ZEUS-02 leading neutron data~\cite{Chekanov:2002pf}. First and foremost, these results demonstrate a good global fit of data taken at different energies Q$^2$ and kinematic ranges of $x$ with our universal set of neutron FFs. 

\begin{figure}[htb]
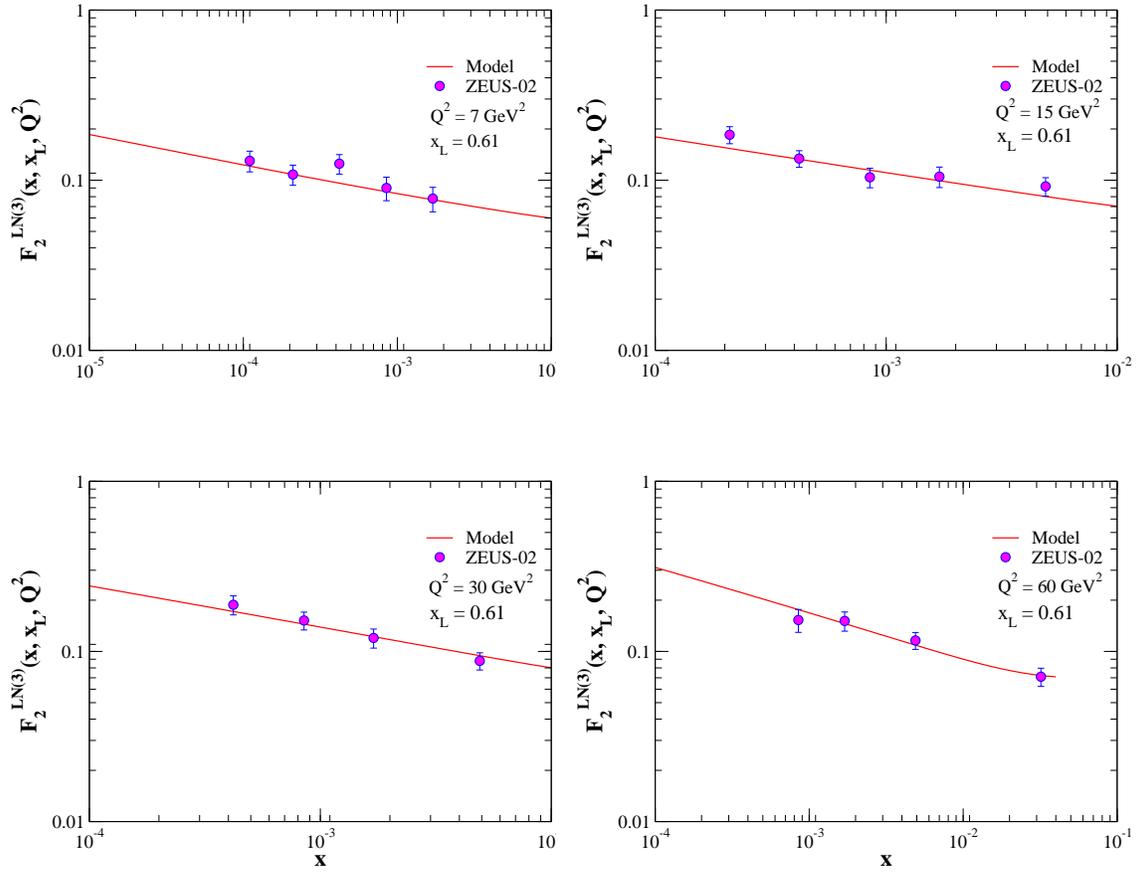

	\begin{center}
		\vspace{0.5cm}
		\resizebox{0.45\textwidth}{!}{\includegraphics{Q2=7-xL=061.eps}}   \vspace{1.cm}  
		\resizebox{0.45\textwidth}{!}{\includegraphics{Q2=15-xL=061.eps}}   
		\resizebox{0.45\textwidth}{!}{\includegraphics{Q2=30-xL=061.eps}}   
		\resizebox{0.45\textwidth}{!}{\includegraphics{Q2=60-xL=061.eps}}   
		\caption{ (Color online) The tagged-neutron structure function $F^{LN(3)}_2$ as a function of $x$ for some selected values of Q$^2$ at fixed value of $x_L = 0.61$. Our theory predictions have been compared with the ZEUS-02 leading neutron data~\cite{Chekanov:2002pf}. }\label{fig:Comparison2}
	\end{center}
\end{figure}

In order to investigate the validity of our QCD analysis and our obtained theory predictions, in Fig.~\ref{fig:Comparison2}, we present a detailed comparison of our results with the ZEUS-02 leading neutron data~\cite{Chekanov:2002pf} for a higher values of $x_L=0.61$.
Overall, one can conclude that our obtained results are in good agreement with all data analyzed, for a wide range of fractional momentum variable $x$ as well as the longitudinal momentum fraction $x_L$.

\clearpage

%
\section{Summary and Conclusions}\label{sec:Summary}

Events with a high energetic leading nucleon carrying a large fraction of the proton beam energy have been observed in $ep$ scattering at HERA~\cite{Chekanov:2002pf,Aaron:2010ab,Rinaldi:2006mf,Chekanov:2008tn}.
In this paper, we have shown that a complete description of semi-inclusive hard processes in perturbabative QCD needs the introduction of new factorizable quantities, known as fracture functions. We have also shown that the fracture functions formalism, in which offers a general theoretical framework
for a QCD-based study of leading baryon physics~\cite{Shoeibi:2017lrl,Ceccopieri:2014rpa,Trentadue:1993ka,Graudenz:1994dq,deFlorian:1997wi,deFlorian:1998rj}, works well in describing the deep inelastic leading-nucleon data measured by the ZEUS collaborations at HERA.
We argued that the fracture functions could open some new possibilities for studying hadron structure and open a new window to predict a variety of hard processes at hadron colliders.
Our analysis is based on the fracture function approach, in which in this framework, semi-inclusive cross sections of leading nucleon production may be written in terms of perturbatively calculable hard-scattering coefficient functions convoluted with appropriate sets of non-perturbative but universal input nucleon FFs constrained by data.
Such a picture was presented here, together with the results of a NLO QCD global analysis of leading nucleon data where it was implemented.
We discuss novel aspect of the methodology used in the present analysis, namely an optimized parametrization of nucleon FFs. We compare {\tt STKJ17} nucleon FFs set to available leading neutron and leading proton data finding in general a reasonable agreement.
We have shown that the partial separation of the nucleon FFs for the various quark flavors has been possible because of the existence of tagged-neutron structure function data from ZEUS-02, and cross section ratio from ZEUS-06 and ZEUS-09 experiments as well as a proposed parametrizations. To further decompose the quark and antiquark neutron FFs, and better constrain the gluon density, additional information will be needed from leading neutron and proton productions in proton-proton collisions.
We obtained a relatively good overall description of the leading neutron and proton productions data at both low and high values of $x$, $x_L$ and Q$^2$.

%
\section*{Acknowledgments}

The authors are especially grateful Luca Trentadue, Garry Levman and Federico Alberto Ceccopieri for many useful discussions and comments.
Hamzeh Khanpour is indebted the University of Science and Technology of Mazandaran and the School of Particles and Accelerators, Institute for Research in Fundamental Sciences (IPM), to support financially this project. Fatemeh Taghavi-Shahri and Kurosh Javidan also acknowledge Ferdowsi University of Mashhad. This work is supported by Ferdosi University of Mashhad under grant number 3/40886 (04/03/1395).


%
%

%

\end{document}